\RequirePackage{ifpdf}
\ifpdf 
\documentclass[pdftex]{sigma}
\else
\documentclass{sigma}
\fi

\usepackage{bm}

\newcommand{\Z}{\mathbb{Z}}
\newcommand{\N}{\mathbb{N}}

\newcommand{\C}{\mathbb{C}}

\newcommand{\D}{\partial}
\newcommand{\op}{\oplus}
\newcommand{\bu}{\bm{u}}
\newcommand{\bv}{\bm{v}}
\newcommand{\blambda}{\bm{\lambda}}

\begin{document}

\allowdisplaybreaks

\renewcommand{\thefootnote}{$\star$}

\renewcommand{\PaperNumber}{025}

\FirstPageHeading

\ShortArticleName{Higher Genus Abelian Functions Associated with Cyclic Trigonal Curves}

\ArticleName{Higher Genus Abelian Functions Associated\\ with Cyclic Trigonal Curves\footnote{This
paper is a contribution to the Proceedings of the Eighth
International Conference ``Symmetry in Nonlinear Mathematical
Physics'' (June 21--27, 2009, Kyiv, Ukraine). The full collection
is available at
\href{http://www.emis.de/journals/SIGMA/symmetry2009.html}{http://www.emis.de/journals/SIGMA/symmetry2009.html}}}

\Author{Matthew ENGLAND}

\AuthorNameForHeading{M. England}

\Address{Department of Mathematics, University of Glasgow, UK}
\Email{\href{mailto:m.england@maths.gla.ac.uk}{m.england@maths.gla.ac.uk}}
\URLaddress{\url{http://www.maths.gla.ac.uk/~mengland/}}

\ArticleDates{Received December 31, 2009, in f\/inal form March 19, 2010;  Published online March 24, 2010}

\Abstract{We develop the theory of Abelian functions associated with cyclic trigonal curves by considering two new cases.  We investigate curves of genus six and seven and consider whether it is the trigonal nature or the genus which dictates certain areas of the theory.  We present solutions to the Jacobi inversion problem, sets of relations between the Abelian function, links to the Boussinesq equation and a new addition formula.}


\Classification{33E05; 14H40; 14H42}

\section{Introduction} \label{SEC_Intro}

Recent times have seen a revival of interest in the theory of Abelian functions associated with algebraic curves.  An Abelian function may be def\/ined as one that has multiple independent periods, in this case derived from the periodicity property of an underlying algebraic curve.  The topic can be dated back to the Weierstrass theory of elliptic functions which we use as a model.

Let $\wp(u)$ be the \textit{Weierstrass $\wp$-function} which, as an elliptic function, has two complex periods~$\omega_1$,~$\omega_2$:
\begin{equation} \label{eq:Intro_period}
\wp(u + \omega_1) = \wp(u + \omega_2) = \wp(u), \qquad \mbox{for all} \ \ u \in \C.
\end{equation}
Elliptic functions have been the subject of much study since their discovery in the 1800s.  The $\wp$-function is a particularly important example which has the simplest possible pole structure for an elliptic function.
The $\wp$-function satisf\/ies a number of interesting properties.  For example, it can be used to parametrise an elliptic curve
\begin{equation*} 
y^2 = 4x^3 - g_2x - g_3,
\end{equation*}
where $g_2$ and $g_3$ are constants.  It also satisf\/ies the following well-known dif\/ferential equations
\begin{gather}
\big(\wp'(u)\big)^2  = 4\wp(u)^3 - g_2\wp(u) - g_3,            \label{eq:Intro_elliptic_diff1}     \\
           \wp''(u)  = 6\wp(u)^2 -  \tfrac{1}{2}g_2.  \label{eq:Intro_elliptic_diff2}
\end{gather}
Weierstrass introduced an auxiliary function, $\sigma(u)$, in his theory which satisf\/ied
\begin{gather}
\wp(u)  = - \frac{d^2}{d u^2} \log \big[ \sigma(u) \big].  \label{eq:Intro_elliptic_ps}
\end{gather}
This $\sigma$-function plays a crucial role in the generalisation and in applications of the theory.  It satisf\/ies the following two term addition formula
\begin{gather}
- \frac{\sigma(u+v)\sigma(u-v)}{\sigma(u)^2\sigma(v)^2} = \wp(u) - \wp(v). \label{eq:Intro_elliptic_add}
\end{gather}
Taking logarithmic derivatives of this will give the standard addition formula for the $\wp$-function.  In this document we present generalisations of equations (\ref{eq:Intro_elliptic_diff1})--(\ref{eq:Intro_elliptic_add}) for new sets of functions.

Klein developed an approach to generalise the Weierstrass $\wp$-function as described in Baker's classic texts \cite{ba97} and \cite{ba07}.  This approach has motivated the general def\/initions in \cite{bel97} and \cite{eel00} of what we now call \emph{Kleinian $\wp$-functions}.  These are def\/ined using the properties of certain algebraic curves.

\begin{definition} \label{def:HG_general_curves}
For two coprime integers $(n,s)$ with $s>n$ we def\/ine a \textit{cyclic $(n,s)$-curve} as a~non-singular algebraic curve
\begin{gather*} 
f(x,y)=0, \qquad f(x,y) = y^n - (x^s + \lambda_{s-1}x^{s-1} + \dots + \lambda_{1}x + \lambda_0 ).
\end{gather*}
Here $x$, $y$ are complex variables while the $\lambda_j$ are a set of curve constants.  (Note that in the literature the word cyclic is sometimes replaced by \emph{strictly} or \emph{purely}.)
\end{definition}
In each case the genus of the curve is given by $\frac{1}{2}(n-1)(s-1)$ and the associated functions become multivariate with $g$ variables, $\bm{u} = (u_1, \dots, u_g)$.  We def\/ine the period matrices,
$\omega'$, $\omega''$ associated with this curve as standard and denote the period lattice formed from~$\omega'$,~$\omega''$ by~$\Lambda$.  These are the points
\[
\Lambda = \big\{ \omega'\bm{m} + \omega''\bm{n}, \  \bm{m},\bm{n} \in \Z^g \big\}.
\]
We consider functions that are periodic with respect to these matrices, or equivalently, invariant under translations by this period lattice.

\begin{definition} \label{def:HG_Abelian}
Let $\mathfrak{M}(\bu)$ be a meromorphic function of $\bu \in \C^g$.  Then $\mathfrak{M}$ is an \textit{Abelian function associated with $C$} if
\begin{equation} \label{eq:HG_Abelian}
\mathfrak{M}(\bu + \omega' \bm{m} + \omega'' \bm{n}) = \mathfrak{M}(\bu),
\end{equation}
for all integer column vectors $\bm{m},\bm{n} \in \Z^g$.
\end{definition}
Note the comparison with equation (\ref{eq:Intro_period}) and that the period matrices play the role of the scalar periods in the elliptic case.

We can def\/ine generalisations of the Weierstrass functions where the periodicity conditions are with respect to these matrices.  We will follow the notation and def\/initions used in~\cite{MEe09} and so only give a brief recap here.   We start by generalising the Weierstrass $\sigma$-function using the properties of the multivariate $\theta$-functions.  The core properties of $\sigma(\bu)$ is that it is entire and that is satisf\/ies a quasi-periodicity condition.  It also has def\/inite parity, dependent on $(n,s)$, with respect to the change of variables $\bu \to [-1]\bu$.  The only zeros of the $\sigma$-function are of order one and occur on the theta-divisor of the Jacobian.

We next def\/ine $\wp$-functions using an analogy of equation (\ref{eq:Intro_elliptic_ps}).
Since there is more than one variable we need to be clear which we dif\/ferentiate with respect to.  We actually def\/ine multiple $\wp$-functions and introduce the following new notation.

\begin{definition} \label{def:HG_2ip}
Def\/ine \textit{$n$-index Kleinian $\wp$-functions}, (where $n\geq2$) as
\begin{eqnarray*}
\wp_{i_1,i_2,\dots,i_n}(\bu) = - \frac{\D}{\D u_{i_1}} \frac{\D}{\D u_{i_2}}\cdots \frac{\D}{\D u_{i_n}} \log \big[ \sigma(\bu) \big],
\qquad i_1 \leq \cdots \leq i_n \in \{1,\dots,g\}.
\end{eqnarray*}
\end{definition}

The $n$-index $\wp$-functions are meromorphic with poles of order $n$ when $\sigma(\bu)=0$.  We can check that they satisfy equation (\ref{eq:HG_Abelian}) and hence they are Abelian.   The $n$-index $\wp$-functions have def\/inite parity with respect to the change of variables $\bu \to [-1]\bu$ and are odd if $n$ is odd and even if $n$ is even.

When the $(n,s)$-curve is chosen to be the classic elliptic curve then the Kleinian $\sigma$-function coincides with the classic $\sigma$-function and the sole 2-index $\wp$-function coincides with the Weierstrass $\wp$-function.  The only dif\/ference would be the notation with
\begin{gather*}
\wp_{11}(\bu) \equiv \wp(u), \qquad \wp_{111}(\bu) \equiv \wp'(u), \qquad \wp_{1111}(\bu) \equiv \wp''(u).
\end{gather*}
As the genus of the curve increases so do the number of associated $\wp$-functions.
In general, the number of $r$-index $\wp$-functions associated with a genus $g$ curve is
$( g + r - 1)!/[r!(g - 1)!]$.

Those curves with $n=2$ are def\/ined to be \emph{hyperelliptic} and the original functions of Klein and Baker were associated with the simplest of these, the (2,5)-curve which has genus two.  Klein considered hyperelliptic curves of arbitrary genus and Baker also constructed examples using a genus three hyperelliptic curve in~\cite{ba03}.  However a full theory for the functions associated to an arbitrary hyperelliptic curve did not follow until the 1990s when Buchstaber, Enolskii and Leykin published~\cite{bel97}.

In the last few years a good deal of progress has been made on the theory of Abelian functions associated to those $(n,s)$ curve with $n=3$, which we label \textit{trigonal curves}.  In \cite{bel00}, the authors of~\cite{bel97} furthered their methods to the trigonal cases, obtaining realisations of the Jacobian variety and some key dif\/ferential equations between the functions.

Over recent years several groups of authors have begun to investigate other aspects of the theory of Abelian functions associated to trigonal curves.  In particular, the two canonical cases of the (3,4)- and (3,5)-curves have been examined in \cite{eemop07} and~\cite{bego08} respectively.   Both papers explicitly construct the dif\/ferentials on the curve, solve the Jacobi inversion problem and obtains sets of dif\/ferential equations between the $\wp$-functions.
The class of $(n,s)$-curves with $n=4$ are def\/ined as \textit{tetragonal curves} and have been recently considered for the f\/irst time in \cite{MEe09} and~\cite{MEg09}.

In this paper we present new results for two higher genus trigonal curves.  The aim was to give further examples of results to build towards a general trigonal case and to investigate the question of whether it is the classif\/ication of the curve, (value of $n$), or the genus that dictates the behaviour.  The results here build on the work in \cite{eemop07} and \cite{bego08} but have used techniques inspired by the problems encountered when working with the tetragonal curve of genus six in \cite{MEe09}.

In Section \ref{SEC_37} we introduce the (3,7)-curve and in Section \ref{SEC_KF} we use a key result to solve the Jacobi inversion problem for this case.  We then proceed to construct a series expansion for the $\sigma$-function in Section \ref{SEC_sig}, including details of the algorithm to calculate the Schur--Weierstrass polynomials in Appendix \ref{APP_SW}.  Relations between the Abelian functions are derived in Section \ref{SEC_Rel} and in Section \ref{SEC_ADD} we construct an addition formula, presenting the full formula in Appendix \ref{APP_add}.  In Section \ref{SEC_38} we give some details relating to the (3,8)-curve and in Appendix \ref{APP_JIP} we discuss the Jacobi inversion problem for curves of even higher genus.  Note that details of calculations which were considered too unwieldy for inclusion have been made available online at \cite{37web}.

\section{The cyclic trigonal curve of genus six} \label{SEC_37}

We will work with the cyclic (3,7)-curve which, from Def\/inition~\ref{def:HG_general_curves}, is the non-singular curve
\begin{gather*}
y^3 = x^7 + \lambda_6x^6 + \lambda_5x^5 + \lambda_4 x^4
+ \lambda_3 x^3 + \lambda_2 x^2+\lambda_1 x + \lambda_0.  
\end{gather*}
Note that this curve has genus $g=6$ which is the same as the cyclic (4,5)-curve investigated in~\cite{MEe09}.
We denote the Riemann surface def\/ined by an $(n,s)$-curve with $C$ and start by deriving dif\/ferentials on this surface.  We can derive the basis of holomorphic dif\/ferentials using the Weierstrass gap-sequence (see for example \cite{bg06})
\begin{gather}
\bm{du} = (du_1, \dots ,du_6), \qquad \mbox{where} \quad du_i(x,y) = \frac{g_i(x,y)}{3y^2}dx,\label{eq:37_holodiff}
\end{gather}
with
\begin{alignat*}{4}
& g_1(x,y) = 1,  \qquad &&   g_2(x,y) = x, \qquad  && g_3(x,y) = x^2, &\nonumber\\
 & g_4(x,y) = y,   \qquad && g_5(x,y) =x^3,  \qquad && g_6(x,y) = xy.&
\end{alignat*}
Although the dimension of the space is the same as the (4,5)-case, the basis of dif\/ferentials is dif\/ferent.  We usually denote a point in $\C^6$ by $\bu$, a row vector with coordinates $(u_1, \dots ,u_6)$.  Any point $\bu \in \C^6$ may be written as follows, where the $P_i$ are a set of variable points upon $C$
\begin{equation} \label{eq:HG_CgfromsymC}
\bu = \sum_{i=1}^6 \int_{\infty}^{P_i} \bm{du}.
\end{equation}

We also work with a particular 2-form known as the \emph{fundamental differential of the second kind}.  This dif\/ferential is symmetric with a pole down the diagonal and a particular series expansion.  (See \cite{bg06} for details of the explicit construction.)  It is used in Klein's formula for the $\wp$-functions discussed in the next section.  The fundamental dif\/ferential may be written as
\begin{equation} \label{eq:FundDiff}
\Omega \big( (x,y), (z,w) \big) = \frac{ F \big( (x,y), (z,w) \big) dx dz}{(x-z)^2f_y(x,y)f_w(z,w)},
\end{equation}
where $F$ is the following polynomial of its variables
\begin{gather}
F \big( (x,y), (z,w) \big)
= 2\lambda_{1}yz + \lambda_{2}x^2w + 2\lambda_{1}wx + 3w\lambda_{0}
+ 3x\lambda_{3}yz^2 - yz^4\lambda_{4} \nonumber \\
\qquad {}+ 2x^2yz^3\lambda_{5} + 4xyz^3\lambda_{4} + yz^4x\lambda_{5} + \lambda_{2}yz^2 + 3yz^4x^2\lambda_{6}
+ 3x^4w\lambda_{6}z^2 + yz^4x^3  \nonumber \\
\qquad {}+ 2x^3w\lambda_{5}z^2 + 2x^5wz^2 + 2x^2yz^5 + x^4wz^3
+ 3y^2w^2 + \lambda_{1}yx + \lambda_{1}wz + 3y\lambda_{0}  \nonumber \\
\qquad {} + \lambda_{5}x^4wz - \lambda_{4}x^4w + 2\lambda_{2}xwz + 2\lambda_{2}xyz + 3\lambda_{3}x^2wz
+ 4\lambda_{4}x^3wz. \label{eq:37_fundF}
\end{gather}
Through this construction we set a basis for the dif\/ferentials of the second kind,
\begin{gather*}
 \bm{dr} = (dr_1,\dots,dr_6), \qquad \mbox{where} \quad dr_j(x,y) = \frac{h_j(x,y)}{3y^2}dx,
\end{gather*}
with
\begin{gather*}
h_1(x,y) = y\big( 9x^4\lambda_6 + 5x^2\lambda_4 + 11x^5 + \lambda_2 + 3x\lambda_3 + 7x^3\lambda_5 \big), \\
h_2(x,y) = xy\big( 4x\lambda_5 + 6x^2\lambda_6 + 8x^3 + 2\lambda_4 \big), \\
h_3(x,y) = y\big( 5x^3 - \lambda_4 + x\lambda_5 + 3x^2\lambda_6 \big),
\qquad h_5(x,y) = 2x^2y, \\
h_4(x,y) = x^3\big( 3x\lambda_6 + 4x^2 + 2\lambda_5 \big),
\qquad h_6(x,y) = x^4.
\end{gather*}

We def\/ine the period matrices by taking integrals of $\bm{du}$ around cycles on the surface.
The surface $C$ is associated with an Abelian variety of dimension $g$, called the Jacobian of the curve, $J = \C^g / \Lambda$.  We def\/ine the \textit{Abel map} as below to move between the curve and the Jacobian{\samepage
\begin{gather}
\mathfrak{A}:  \ \mbox{Sym}^k(C) ,\to     J, \nonumber \\
\phantom{\mathfrak{A}:} \ \ (P_1,\dots,P_k)   ,\mapsto \left( \int_{\infty}^{P_1} \bm{du} + \dots + \int_{\infty}^{P_k} \bm{du} \right) \pmod{\Lambda}. \label{eq:HG_AbelMap}
\end{gather}
Here the $P_i$ are points upon the curve $C$.}

We then def\/ine the $\wp$-functions as in general.  Since the genus is six we will, with respect to notation, be working with the same Abelian functions as in the (4,5)-case.  However, we f\/ind the results dif\/fer which is seen most clearly by the dif\/ferent weights of the functions.

Within the theory of these functions we can def\/ine weights such that every equation is homogeneous.  In general they can be predicted by the values of $(n,s)$ and the Weierstrass gap sequence these def\/ine.  In the (3,7)-case these weights are as follows
\begin{center}
\begin{tabular}{|c|c|c|c|c|c|c|c|c|c|c|c|c|}\hline
                & $\phantom{+}x$  & $\phantom{+}y$   &  $\phantom{+}\lambda_6$ & $\phantom{+}\lambda_5$ & $\phantom{+}\lambda_4$ & $\phantom{+}\lambda_3$ & $\phantom{+}\lambda_2$ & $\phantom{+}\lambda_1$ & $\phantom{+}\lambda_0$    \\ \hline
{Weight} & $-3$ & $-7$  &  $-3$ & $-6$ & $-9$  & $-12$        & $-15$       & $-18$       & $-21$          \\ \hline
\end{tabular}

\medskip

\begin{tabular}{|c|c|c|c|c|c|c|c|c|c|c|c|c|}\hline
                & $u_1$ & $u_2$  &  $u_3$ & $u_4$ & $u_5$ & $u_6$ & $\sigma(\bu)$ \\ \hline
{Weight} & $+11$ & $+8$   &  $+5$  & $+4$  & $+2$  & $+1$  & $+ 16$        \\ \hline
\end{tabular}
\end{center}
Note that the weights of the $\wp$-functions are equal to the negative of the sum of the weights of their indices.  The weight properties were derived and justif\/ied in general throughout \cite{MEe09}, although they were known and used previously.  They are often referred to as \textit{Sato weights}.

\section{Expanding the Kleinian formula} \label{SEC_KF}

In this section we work with the Kleinian formula, (Theorem~3.4 in \cite{eel00}), which links the $\wp$-functions with a point on the curve.  Let $\{(x_1,y_1),\dots,(x_6,y_6)\} \in C^6$ be an arbitrary set of distinct points and $(z,w)$ any point of this set.  Then for an arbitrary point $(x,y)$ and base point~$\infty$ on $C$ we have
\begin{equation} \label{eq:37_Klein}
\sum_{i,j=1}^6 \wp_{ij} \left( \int_{\infty}^t \bm{du} - \sum_{k=1}^6 \int_{\infty}^{x_k} \bm{du} \right)
g_i(x,y) g_j(z,w)
= \frac{F\big((x,y),(z,w)\big)}{(x-z)^2}.
\end{equation}
Here $g_i$ is the numerator of $du_i$, as given in equation (\ref{eq:37_holodiff}), and $F$ is the symmetric function given in equation (\ref{eq:37_fundF}) as the numerator of the fundamental dif\/ferential of the second kind.

We will extract information from this formula by expanding it as a series and setting each coef\/f\/icient to zero.  We use expansions of the variables in  $\xi$, the local parameter in the neighborhood of the inf\/inite point, $\infty \in C$.  This is
def\/ined by $\xi = x^{-\frac{1}{3}}$ and the following expansions can be easily derived from the def\/initions of the variables involved
\begin{gather*}
x = \frac{1}{\xi^{3}}, \qquad
y = \frac{1}{\xi^7} + \left(\frac{\lambda_6}{3}\right)\frac{1}{\xi^4}
+ \left( \frac{\lambda_5}{3} - \frac{\lambda_6^2}{9} \right)\frac{1}{\xi}
+ \left( \frac{\lambda_4}{3} - \frac{2\lambda_6\lambda_5}{9} + \frac{5\lambda_6^3}{81} \right) \xi^2
+ O\big( \xi^5 \big),\nonumber
\\
u_1 = -\tfrac{1}{11}\xi^{11} + O(\xi^{14}), \qquad u_3= -\tfrac{1}{5}\xi^5 + O(\xi^{8}),
\qquad u_5 = -\tfrac{1}{2}\xi^2 + O(\xi^5), \nonumber\\
u_2 = -\tfrac{1}{8}\xi^{8} + O(\xi^{11}),   \qquad u_4= -\tfrac{1}{4}\xi^4 + O(\xi^7),
\quad u_6 = -\xi + O(\xi^4).
\end{gather*}
We substitute these expansions into the Kleinian formula to obtain a series in $\xi$, equal to zero.  The coef\/f\/icients give equations in the variables $(z,w)$ and the $\wp$-functions, starting with the three below
\begin{gather}
0 = \rho_1 =
-z^4 + \wp_{56}z^3 + \wp_{36}z^2 + \big(\wp_{66}w + \wp_{26}\big)z + \wp_{46}w + \wp_{16}, \label{eq:37_pp1} \\
0  = \rho_2 =
\big(\wp_{55} -\wp_{566} \big)z^3 + \big(\wp_{35} -2w - \wp_{366} \big)z^2
+ \big(\wp_{25} -\wp_{266} + \wp_{56}w   \nonumber \\
\phantom{0  = \rho_2 =}{} - \wp_{666}w \big)z
+ \wp_{15} - \wp_{466}w - \wp_{166} + \wp_{45}w, \nonumber\\ 
0  = \rho_3 =  \big( \tfrac{1}{2}\wp_{5666} - \tfrac{3}{2}\wp_{556} \big)z^3
+ \big( \tfrac{1}{2}\wp_{3666} - \tfrac{3}{2}\wp_{356} \big)z^2
+ \big( \tfrac{1}{2}\wp_{2666} - \tfrac{3}{2}\wp_{256}    \nonumber \\
\phantom{0  = \rho_3 =}{} + (\tfrac{1}{2}\wp_{6666} - \tfrac{3}{2}\wp_{566} )w  \big)z
- \tfrac{3}{2}\wp_{156} - 3w^2 + \tfrac{1}{2}\wp_{1666}
+ \big( - \tfrac{3}{2}\wp_{456} + \tfrac{1}{2}\wp_{4666} \big)w,  \nonumber \\
 \vdots
 \nonumber
\end{gather}
They are valid for any $\bu \in \C^6$, with $(z,w)$ one of the points on $C$ that may be used to represent~$\bu$ in equation (\ref{eq:HG_CgfromsymC}).  They are presented in ascending order (as the coef\/f\/icients of $\xi$) and have been calculated explicitly (using Maple), up to $\rho_{10}$.  They get increasingly large in size and can be found at \cite{37web}.  We aim to manipulate these relations to eliminate $(z,w)$.

We start by taking resultants of pairs of equations, eliminating the variable $w$ by choice, and denoting the resultant of $\rho_i$ and $\rho_j$ by $\rho_{i,j}$.  The f\/irst of these, $\rho_{1,2}$, is presented later in equation~(\ref{eq:37_rho12}) as part of the main theorem in this section.  The other $\rho_{i,j}$ get increasingly large and so we instead present Table~\ref{tab:37_rhoij} which details the number of terms they contain once expanded and their degree in $z$.
In general, the higher the $\rho_i$ involved in the resultant, the more terms the resultant has.  There were only two polynomials found with degree six in $z$, $\rho_{12}$ and $\rho_{14}$, with all the others having higher degree.  In the (4,5)-case the corresponding polynomials had many more terms and all had degree in $z$ of at least seven.  These calculations have far more in common with the lower genus trigonal cases than the tetragonal case of the same genus.
\begin{table}[ht]
\caption{The polynomials $\rho_{i,j}$.}  \label{tab:37_rhoij}
\begin{center}
\begin{tabular}{ l | l | c c || c l | l | c }
Res           & $\#$ terms & degree &        &        & Res           & $\#$ terms & degree \\
    \hline
$\rho_{1,2}$  & 40         & 6      & \qquad & \quad \qquad & $\rho_{3,4}$  & 2025       & 10 \\
$\rho_{1,3}$  & 79         & 8      & \qquad & \quad \qquad & $\rho_{3,5}$  & 4188       & 9  \\
$\rho_{1,4}$  & 77         & 6      & \qquad & \quad \qquad & $\rho_{3,6}$  & 4333       & 8  \\
$\rho_{1,5}$  & 154        & 7      & \qquad & \quad \qquad & $\rho_{3,7}$  & 19043      & 10 \\
$\rho_{1,6}$  & 344        & 8      & \qquad & \quad \qquad & $\rho_{3,8}$  & 28422      & 10 \\
$\rho_{1,7}$  & 290        & 7      & \qquad & \quad \qquad & $\rho_{3,9}$  & 44409      & 10 \\
$\rho_{1,8}$  & 412        & 7      & \qquad & \quad \qquad & $\rho_{4,5}$  & 793        & 8  \\
$\rho_{1,9}$  & 1055       & 8      & \qquad & \quad \qquad & $\rho_{4,6}$  & 8315       & 10 \\
$\rho_{2,3}$  & 307        & 7      & \qquad & \quad \qquad & $\rho_{4,7}$  & 1183       & 8  \\
$\rho_{2,4}$  & 219        & 7      & \qquad & \quad \qquad & $\rho_{4,8}$  & 2112       & 8  \\
$\rho_{2,5}$  & 226        & 6      & \qquad & \quad \qquad & $\rho_{4,9}$  & 24569      & 10 \\
$\rho_{2,6}$  & 1468       & 8      & \qquad & \quad \qquad & $\rho_{5,6}$  & 18356      & 10 \\
$\rho_{2,7}$  & 712        & 7      & \qquad & \quad \qquad & $\rho_{5,7}$  & 2535       & 8  \\
$\rho_{2,8}$  & 737        & 7      & \qquad & \quad \qquad & $\rho_{5,8}$  & 2316       & 8  \\
$\rho_{2,9}$  & 4536       & 9      & \qquad & \quad \qquad & $\rho_{5,9}$  & 54384      & 11
\end{tabular}
\end{center}
\end{table}

We need to reduce these equations to degree f\/ive ($= g-1$) in $z$ since such equations must be satisf\/ied by all six variables and so are identically zero.  We select $\rho_{1,2}$ as it is the smallest polynomial of degree six, and rearrange it to give an equation for $z^6$.  Note that this equation is polynomial since the coef\/f\/icient of $z^6$ in $\rho_{1,2}$ was a constant
\begin{gather*}
z^6 =   z^5\big( \tfrac{3}{2}\wp_{56} - \tfrac{1}{2}\wp_{666} \big) + z^4\big(\tfrac{1}{2}\wp_{55}\wp_{66} + \cdots.
\end{gather*}
We may now take any of the other $\rho_{i,j}$ and repeatedly substitute for $z^6$ until we have a polynomial of degree f\/ive in $z$.  We can then set the coef\/f\/icients to zero leaving us with six polynomial equations between the $\wp$-functions.  Finally, recalling that the $\wp$-functions have def\/inite parity, we can separate each of these six relations into their odd and even parts.
In the (4,5)-case every polynomial needed at least two rounds of substitution to reach this stage, with the corresponding equations for $z^6$ and $z^7$ far more complicated.  In this case we need only substitute once into~$\rho_{1,4}$ and so the polynomials achieved here are far simpler.  They start with
\begin{gather*}
0 =
\big( 2\wp_{666} -3\lambda_{{6}} -6\wp_{56}  \big) \wp_{66}
+\wp_{5666} -3\wp_{46}, \\
0 =
-\tfrac{1}{2}\wp_{556} -\tfrac{1}{6}\wp_{66666},  \\
0  =
\big( 2\wp_{566} -2\wp_{55}  \big)  \wp_{66}^{2}
+ \big( \wp_{5566} -\wp_{45} -2\lambda_{{5}}
-4\wp_{36} +2\wp_{466} \\
\phantom{0=}{} -2\wp_{56} \wp_{666}
+2\wp_{56}^{2} \big) \wp_{66}
+\wp_{44} +\wp_{4566} -\wp_{56} \wp_{5666}
-\wp_{56} \wp_{46} -3\wp_{46} \lambda_{{6}}, \\
0  =
- \big( \tfrac{1}{6}\wp_{56666} + \tfrac{1}{2}\wp_{555}  \big) \wp_{66}
-\tfrac{1}{6}\wp_{46666} +\tfrac{1}{2}\wp_{56} \wp_{556}
+\tfrac{1}{6}\wp_{56} \wp_{66666} -\tfrac{1}{2}\wp_{455},  \\
\vdots
\end{gather*}
The other relations in this case all involve two rounds of substitution for $z^6$ and get increasingly long and complex, involving higher index $\wp$-functions.  Although the relations we achieve are far simpler than those in the (4,5)-case, there are not enough simple relations to manipulate in order to f\/ind the desired dif\/ferential equations between the $\wp$-functions.  Hence we proceed, in Sections \ref{SEC_sig} and \ref{SEC_Rel}, to use the tools developed for the (4,5)-curve to f\/ind these relations.

However, as predicted by \cite{bel00}, we may still use the relations from the Kleinian formula to f\/ind the solution of the Jacobi inversion problem.  Recall that the Jacobi inversion problem is, given a point $\bm{u} \in J$, to f\/ind the preimage of this point under the Abel map, (\ref{eq:HG_AbelMap}).

\begin{theorem} \label{thm:37_JIP}
Suppose we are given $\{u_1, \dots, u_6\} = \bm{u} \in J$.  Then we could solve the Jacobi inversion problem explicitly as follows.  We first consider the polynomial $\rho_{1,2}$ which has degree six in $z$ and is equal to zero,
\begin{gather}
\rho_{1,2} = - 2z^6 + ( 3\wp_{56} - \wp_{666} )z^5
+ (\wp_{55}\wp_{66} + 2\wp_{36} - \wp_{56}^2 + \wp_{45} - \wp_{566}\wp_{66} + \wp_{56}\wp_{666} \nonumber \\
\phantom{\rho_{1,2} =}{} - \wp_{466})z^4 + (2\wp_{26} - \wp_{56}\wp_{45} + \wp_{35}\wp_{66} - \wp_{566}\wp_{46}
+ \wp_{55}\wp_{46} - \wp_{366}\wp_{66} + \wp_{56}\wp_{466} \nonumber \\
\phantom{\rho_{1,2} =}{} - \wp_{36}\wp_{56} + \wp_{36}\wp_{666})z^3 + (2\wp_{16} + \wp_{35}\wp_{46} - \wp_{266}\wp_{66}
- \wp_{366}\wp_{46} + \wp_{26}\wp_{666} - \wp_{26}\wp_{56} \nonumber \\
\phantom{\rho_{1,2} =}{}
 + \wp_{25}\wp_{66} - \wp_{36}\wp_{45} + \wp_{36}\wp_{466})z^2
+ ( \wp_{16}\wp_{666} - \wp_{266}\wp_{46} - \wp_{16}\wp_{56} - \wp_{26}\wp_{45} + \wp_{15}\wp_{66} \nonumber \\
\phantom{\rho_{1,2} =}{} - \wp_{166}\wp_{66} + \wp_{25}\wp_{46} + \wp_{26}\wp_{466})z - \wp_{16}\wp_{45} - \wp_{166}\wp_{46}
+ \wp_{15}\wp_{46} + \wp_{16}\wp_{466}. \label{eq:37_rho12}
\end{gather}
Denote by $(z_1,\dots,z_6)$ the six zeros of this polynomial, which will be expressions in $\wp$-functions that should be evaluated for the given $\bm{u}$.  Next recall equation \eqref{eq:37_pp1} which is the following equation of degree one in $w$
\[
0 = -z^4 + \wp_{56}z^3 + \wp_{36}z^2 + \big(\wp_{66}w + \wp_{26}\big)z + \wp_{46}w + \wp_{16}.
\]
We can substitute each $z_i$ into this equation in turn and solve to find the corresponding $w_i$.
Therefore the set of points $\{(z_1, w_1), \dots, (z_6,w_6)\}$ on the curve $C$ which are the Abel preimage of $\bm{u}$ have been identified.
\end{theorem}

Note that since $\rho_{1,2}$ is degree six in $z$ and equal to zero, the coef\/f\/icients in equation (\ref{eq:37_rho12}) will be elementary symmetric functions in the roots of the equation.  Hence they give expressions for the elementary symmetric functions in the coordinates of the divisor of six points on $C$.

Note also that although Theorem \ref{thm:37_JIP} only required the f\/irst few relations derived from the Kleinian formula, many of the other relations are used implicitly in the construction of the $\sigma$-function expansion in the next section.

\section[The $\sigma$-function expansion]{The $\boldsymbol{\sigma}$-function expansion} \label{SEC_sig}

We construct a $\sigma$-function expansion for the cyclic (3,7)-curve using the methods and techniques discussed in detail in \cite{MEe09}.  We start with a statement on the structure of the expansion.

\begin{theorem} \label{thm:37_sigexp}
The function $\sigma(\bu)$ associated with the cyclic {\rm (3,7)}-curve may be expanded about the origin as
\begin{gather*}
\sigma(\textbf{u}) = \sigma(u_1, u_2, u_3, u_4, u_5, u_6) = SW_{3,7}(\bu) +  C_{19}(\bu) + C_{22}(\bu)  + \dots + C_{16 + 3n}(\bu) + \cdots,
\end{gather*}
where $SW_{3,7}$ is the Schur--Weierstrass polynomial and each $C_k$ is a finite, even polynomial composed of products of monomials in $\bm{u} = (u_1,u_2,\dots,u_6)$ of weight $+k$ multiplied by monomials in $\bm{\lambda}=(\lambda_6,\lambda_5,\dots,\lambda_0)$ of weight $16-k$.
\end{theorem}

\begin{proof}
This proof follows the ideas in \cite{eemop07} and \cite{MEe09}.  First we note that the $\sigma$-function associated with the (3,7)-curve is even and then, by Theorem 3(i) in \cite{N08}, we know the expansion will be a~sum of monomials in $\bm{u}$ and $\bm{\lambda}$ with rational coef\/f\/icients.

One of the key properties of the $\sigma$-function is that the part of its expansion without $\blambda$ will be given by a constant multiple of the corresponding Schur--Weierstrass polynomial (see \cite{bel99} or \cite{N08}) and as discussed in \cite{MEe09} we can set this constant to one.
We can then conclude the weight of the expansion to be the weight of $SW_{3,7}$ which is $+16$.  We split the expansion into polynomials with common weight ratios and increasing weight in $\bu$.  The subscripts increase by three since the possible weights of $\blambda$-monomials decrease by three.
\end{proof}

The Schur--Weierstrass polynomial is generated from the integers $(n,s)$ using an easily implemented algorithmic procedure which is summarised in Appendix~\ref{APP_SW}.  In this case
\begin{gather*}
SW_{3,7} =   \tfrac{1}{22528000}u_{6}^{16} + u_{2}^2
+ \tfrac{1}{80}u_{4}u_{6}^6u_{5}^3 + \tfrac{1}{3200}u_{4}u_{6}^{10}u_{5}
+ \tfrac{1}{8}u_{6}^2u_{5}^5u_{4} - \tfrac{1}{2}u_{3}u_{6}^3u_{4}^2
- \tfrac{3}{281600}u_{6}^{12}u_{5}^2  \nonumber \\
\phantom{SW_{3,7} =}{} + \tfrac{1}{2}u_{6}^2u_{5}u_{4}^3
- \tfrac{1}{16}u_{4}^2u_{6}^4u_{5}^2 + u_{6}u_{5}^2u_{1} - \tfrac{1}{2}u_{2}u_{6}^2u_{5}^3 - u_{2}u_{5}^2u_{4} - \tfrac{1}{35200}u_{3}u_{6}^{11}
+ \tfrac{1}{40}u_{2}u_{6}^6u_{5}    \nonumber \\
\phantom{SW_{3,7} =}{}
- \tfrac{1}{20}u_{6}^5u_{1}  - \tfrac{1}{20}u_{4}u_{6}^5u_{3}u_{5} - \tfrac{1}{80}u_{6}^7u_{5}^2u_{3}
+ \tfrac{1}{2}u_{3}^2u_{6}^2u_{5}^2 + 2u_{5}u_{4}u_{3}^2
+ \tfrac{3}{8}u_{5}^4u_{4}^2 - \tfrac{1}{640}u_{4}^2u_{6}^8  \nonumber  \\
\phantom{SW_{3,7} =}{}
+ \tfrac{1}{4}u_{4}u_{6}^4u_{2}
+ \tfrac{1}{40}u_{3}^2u_{6}^6  + \tfrac{1}{8}u_{3}u_{6}^3u_{5}^4 - u_{3}u_{6}u_{5}^3u_{4}
- \tfrac{1}{64}u_{5}^8 + \tfrac{1}{2560}u_{6}^8u_{5}^4
- \tfrac{1}{4}u_{4}^4 - u_{3}u_{1}\\
\phantom{SW_{3,7} =}{}
 - 2u_{2}u_{3}u_{6}u_{5}
- \tfrac{1}{64}u_{5}^6u_{6}^4 + u_{6}u_{3}^3. \nonumber
\end{gather*}
We construct the other $C_{k}$ successively following the methods of \cite{MEe09}.  We can identify the possible terms in each polynomial as the f\/inite number with the desired weight ratio.  We then determine the coef\/f\/icients of each term by ensuring the expansion satisf\/ies known properties of the $\sigma$-function.  These include the vanishing properties, the polynomials derived from the expansion of the Kleinian formula and most ef\/f\/iciently, the equations in the next section found in the construction of basis (\ref{eq:37_BASIS}).  See \cite{MEe09} for full details of this calculation.  The Maple procedures from the (4,5)-case could easily be adapted to these calculations.

We have calculated the $\sigma$-expansion associated with the (3,7)-curve up to and including $C_{49}$.  The polynomials are available online at \cite{37web}.  As in the (4,5)-case we simplify calculations by rewriting procedures to take into account weight simplif\/ications and by using parallel computing implemented with Distributed Maple~\cite{DM}.

The later polynomials are extremely large and represent a signif\/icant amount of computation, in common with the (4,5)-case rather than the lower genus trigonal cases.  In fact these computations are marginally more dif\/f\/icult since the $\blambda$ increase in steps of three instead of four, and so there are a greater number of possible $\blambda$-monomials at each stage.  Hence there are more polynomials $C_k$ with more terms in each.

\section{Relations between the Abelian functions}\label{SEC_Rel}

In this section we derive sets of relations between the Abelian functions associated with the (3,7)-curve.  Note that these relations are presented in weight order as indicated by the bold number in brackets.  We start with the construction of a basis for the space of those Abelian functions with poles of order at most two.  Recall that there are no Abelian functions with simple poles and that an Abelian function with no poles is a constant.  Hence those functions with double poles have the simplest possible pole structure and so are sometimes referred to as \textit{fundamental Abelian functions}.

We f\/ind that for $(n,s)$-curves with genus greater that $2$, the $\wp$-functions are not suf\/f\/icient to span the space.  This problem may be overcome by def\/ining another class of Abelian functions.

\begin{definition} \label{def:45_Qdef}
Def\/ine the operator $\Delta_i$ as below.  This is now known as \textit{Hirota's bilinear operator}, although it was used much earlier by Baker in \cite{ba07}
\[
\Delta_i = \frac{\D}{\D u_i} - \frac{\D}{\D v_i}.
\]
Then an alternative, equivalent def\/inition of the 2-index $\wp$-functions is given by
\begin{equation*} 
\wp_{ij}(\bm{u}) = - \dfrac{1}{2\sigma(\bm{u})^2} \Delta_i\Delta_j \sigma(\bm{u}) \sigma(\bm{v}) \, \Big|_{\bv=\bu},
\qquad i \leq j \in \{1,\dots,g\}.
\end{equation*}
We extend this approach to def\/ine the \textit{${n}$-index ${Q}$-functions}, for $n$ even, by
\begin{equation*} 
Q_{i_1, i_2,\dots,i_n}(\bm{u}) =  \frac{(-1)}{2\sigma(\bm{u})^2} \Delta_{i_1}\Delta_{i_2}\cdots \Delta_{i_n} \sigma(\bm{u}) \sigma(\bm{v}) \, \Big|_{\bv=\bu},
\end{equation*}
where $i_1 \leq \dots  \leq i_n \in \{1,\dots,g\}$.
\end{definition}

The $n$-index $Q$-functions are Abelian functions with poles of order two when $\sigma(\bu)=0$.
They are a generalisation of the $Q$-functions used by Baker.  The 4-index $Q$-functions were introduced in research on the lower genus trigonal curves, (and in the literature are just def\/ined as \textit{$Q$-functions}).  The general def\/inition was developed in \cite{MEe09} as increasing classes are required to deal with cases of increasing genus.
Note that if Def\/inition \ref{def:45_Qdef} were applied with $n$ odd then the resulting function is identically zero.  In \cite{eemop07} it was shown that the 4-index $Q$-functions could be expressed using the Kleinian $\wp$-functions as follows
\begin{gather} \label{eq:45_4iQ}
Q_{ijk\ell} = \wp_{ijk\ell} - 2 \wp_{ij}\wp_{k\ell}-2\wp_{ik}\wp_{j\ell} -2\wp_{i\ell}\wp_{jk}.
\end{gather}
Similarly, in \cite{MEe09} the 6-index $Q$-functions were expressed as
\begin{gather*}
Q_{ijklmn} = \wp_{ijklmn} - 2\Big[
\big(\wp_{ij}\wp_{klmn} + \wp_{ik}\wp_{jlmn} + \wp_{il}\wp_{jkmn} + \wp_{im}\wp_{jkln}
 + \wp_{in}\wp_{jklm}\big) \nonumber \\
\quad
{}+ \big( \wp_{jk}\wp_{ilmn} + \wp_{jl}\wp_{ikmn} + \wp_{jm}\wp_{ikln} + \wp_{jn}\wp_{iklm} \big)
+ \big( \wp_{kl}\wp_{ijmn} + \wp_{km}\wp_{ijln} + \wp_{kn}\wp_{ijlm} \big)  \nonumber \\
\quad {}+ \big( \wp_{lm}\wp_{ijkn}
+ \wp_{ln}\wp_{ijkm} \big) + \wp_{mn}\wp_{ijkl} \Big]
+ 4\Big[ \big( \wp_{ij}\wp_{kl}\wp_{mn} + \wp_{ij}\wp_{km}\wp_{ln} + \wp_{ij}\wp_{kn}\wp_{lm} \big) \nonumber \\
\quad {}+ \big(\wp_{ik}\wp_{jl}\wp_{mn} + \wp_{ik}\wp_{jm}\wp_{ln}
+ \wp_{ik}\wp_{jn}\wp_{lm} \big) + \big(\wp_{il}\wp_{jk}\wp_{mn}  + \wp_{il}\wp_{jm}\wp_{kn}
+ \wp_{il}\wp_{jn}\wp_{km} \big) \nonumber \\
\quad {}+ \big(\wp_{im}\wp_{jk}\wp_{ln}
+ \wp_{im}\wp_{jl}\wp_{kn}  + \wp_{im}\wp_{jn}\wp_{kl} \big) + \big(\wp_{in}\wp_{jk}\wp_{lm}
+ \wp_{in}\wp_{jl}\wp_{km} + \wp_{in}\wp_{jm}\wp_{kl} \big) \Big]. 
\end{gather*}

\begin{theorem} \label{thm:37_basis}
A basis for the space of Abelian functions upon $J$ with poles of order at most two occurring on the theta divisor is given by
\begin{gather} \label{eq:37_BASIS}
\begin{array}{@{}ccccccccccccccc}
    & \C1         &\op& \C\wp_{11}  &\op& \C\wp_{12}  &\op& \C\wp_{13}  &\op& \C\wp_{14}
&\op& \C\wp_{15}  \\
\op & \C\wp_{16}  &\op& \C\wp_{22}  &\op& \C\wp_{23}  &\op& \C\wp_{24}  &\op& \C\wp_{25}
&\op& \C\wp_{26}  \\
\op & \C\wp_{33}  &\op& \C\wp_{34}  &\op& \C\wp_{35}  &\op& \C\wp_{36}  &\op& \C\wp_{44}
&\op& \C\wp_{45}  \\
\op & \C\wp_{46}  &\op& \C\wp_{55}  &\op& \C\wp_{56}  &\op& \C\wp_{66}  &\op& \C Q_{5556}
&\op& \C Q_{5555} \\
\op & \C Q_{4466} &\op& \C Q_{3466} &\op& \C Q_{4456} &\op& \C Q_{4455} &\op& \C Q_{4446}
&\op& \C Q_{3355} \\
\op & \C Q_{3446} &\op& \C Q_{4445} &\op& \C Q_{3346} &\op& \C Q_{3445} &\op& \C Q_{1556}
&\op& \C Q_{2356} \\
\op & \C Q_{1555} &\op& \C Q_{2355} &\op& \C Q_{2446} &\op& \C Q_{2346} &\op& \C Q_{2445}
&\op& \C Q_{3344} \\
\op & \C Q_{2345} &\op& \C Q_{3334} &\op& \C Q_{1355} &\op& \C Q_{1446} &\op& \C Q_{2255}
&\op& \C Q_{2246} \\
\op & \C Q_{2344} &\op& \C Q_{2245} &\op& \C Q_{2334} &\op& \C Q_{1255} &\op& \C Q_{1335}
&\op& \C Q_{1344} \\
\op & \C Q_{2244} &\op& \C Q_{2226} &\op& \C Q_{2234} &\op& \C Q_{1155} &\op& \C Q_{1235}
&\op& \C Q_{1136} \\
\op & \C Q_{1155} &\op& \C Q_{1133} &\op& \C Q_{223466} &\op& \C Q_{113666}. & &
& &
\end{array}
\end{gather}
\end{theorem}
\begin{proof}
The dimension of the space is $2^g=2^6=64$ by the Riemann--Roch theorem for Abelian varieties.
All the selected elements belong to the space and we can easily check their linear independence explicitly in Maple using the $\sigma$-expansion.
\end{proof}

To actually construct the basis we started by including all 21 of the $\wp_{ij}$ since they are all linearly independent.  We then decided which $Q_{ijkl}$ to include by testing at decreasing weight levels to see which could be written as a linear combination.  (Note that these computations were actually performed in tandem with the construction of the $\sigma$-expansion.  Once a new $C_k$ was found three more weight levels of the basis could be examined.  The relations obtained could then be used to construct the next $C_k$ in the expansion.)

Upon examining all the 4-index $Q$-functions, we f\/ind that 62 basis elements have been identif\/ied.  To f\/ind the f\/inal two basis elements we repeat the procedure using the 6-index $Q$-functions.  We f\/ind that all those of weight higher than $-27$ can be expressed as a linear combination of existing basis entries.  However, at weight $-27$ one of the $Q_{ijklmn}$ is required in the basis to express the others, which was the case also at weight $-30$.  The basis requires many Q-functions including some with six indices and so has much more in common with the (4,5)-calculations that the lower genus trigonal work.

\begin{lemma}\label{lem:37_Qcor}
Those $4$-index $Q$-functions not in the basis can be expressed as a linear combination of the basis elements
\begin{gather*}
\bm{(-4)} \quad Q_{6666} = - 3\wp_{55}, \\
\bm{(-5)} \quad Q_{5666} = 3\wp_{46} + 3\lambda_6\wp_{66}, \\
\bm{(-6)} \quad Q_{5566} = 4\wp_{36} - \wp_{45} + 3\lambda_6\wp_{56} + 2\lambda_5, \\
\bm{(-7)} \quad Q_{4666} = 3\lambda_6\wp_{55} - Q_{5556}, \\
\bm{(-8)} \quad Q_{3666} =  -\tfrac{1}{4}Q_{5555} - \tfrac{3}{4}\wp_{44}
+ \tfrac{3}{2}\lambda_6\wp_{46} - \tfrac{3}{4}\lambda_6^2\wp_{66} + 3\lambda_5\wp_{66}, \\
\bm{(-8)} \quad Q_{4566} = -\wp_{44} + 3\lambda_6\wp_{46}, \\
\bm{(-9)} \quad Q_{3566} = 4\wp_{26} - \wp_{34} + 3\lambda_6\wp_{36}, \\
\bm{(-9)} \quad Q_{4556} = 3\wp_{26} - 2\wp_{34} + 3\lambda_6\wp_{45} - 2\lambda_4, \\
\vdots
\end{gather*}
The relations have been calculated down to weight $-37$ and are available at {\rm \cite{37web}}.
There are similar equations for all the $6$-index $Q$-functions, except $Q_{223466}$ and $Q_{113666}$ which were in the basis.  Explicit relations have been calculated down to weight~$-32$.  The first few are given below with the other relations available at~{\rm \cite{37web}}
\begin{gather*}
\bm{(-6)} \quad Q_{666666} = 36\wp_{36} - 45\wp_{45} - 9\lambda_6\wp_{56} - 6\lambda_5, \\
\bm{(-7)} \quad Q_{566666} = -24\wp_{35} + 5Q_{5556} - 24\lambda_6\wp_{55}, \\
\bm{(-8)} \quad Q_{556666} = 12\wp_{44} + Q_{5555} + 24\lambda_6\wp_{46} + 12\lambda_6^2\wp_{66}, \\
\bm{(-9)} \quad Q_{555666} =9\wp_{26} + 18\wp_{34} + 36\lambda_6\wp_{36} - 18\lambda_6\wp_{45}
+ 9\wp_{56}\lambda_6^2 + 6\lambda_6\lambda_5 + 42\lambda_4, \\
\bm{(-9)} \quad Q_{466666} = 15\wp_{26} + 6\wp_{34} - 9\lambda_6\wp_{45} + 6\lambda_4, \\
\vdots
\end{gather*}
\end{lemma}

\begin{proof}
It is clear that such relations must exist.  The explicit dif\/ferential equations were calculated in the construction of the basis.
\end{proof}

Note that we can apply equation (\ref{eq:45_4iQ}) to the f\/irst set of equations in order to derive a~gene\-ra\-lisation of equation (\ref{eq:Intro_elliptic_diff2}), the second dif\/ferential equation from the elliptic case
\begin{gather}
\bm{(-4)} \quad \wp_{6666} = 6\wp_{66}^2 - 3\wp_{55},  \label{Bous} \\
\bm{(-5)} \quad \wp_{5666} = 6\wp_{56}\wp_{66} + 3\wp_{46} + 3\lambda_6\wp_{66},  \label{B2}  \\
\bm{(-6)} \quad \wp_{5566} = 4\wp_{36} - \wp_{45} + 3\lambda_6\wp_{56}
+ 2\lambda_5 + 2\wp_{55}\wp_{66} + 4\wp_{56}^2, \nonumber \\
\vdots
\nonumber
\end{gather}

\begin{remark} \label{rem:37_Bous}
Equation (\ref{Bous}) may be dif\/ferentiated twice with respect to $u_6$ to give the Boussinesq equation for $\wp_{66}$ with $u_6$ playing the space variable and $u_5$ the time variable.  This connection with the Boussinesq equation has been previously established in \cite{eel00, bel00, eemop07} and \cite{bego08}, and is discussed further in Remark~\ref{rem:38_Bous}.
\end{remark}

There are also relations to generalisation equation (\ref{eq:Intro_elliptic_diff1}) from the elliptic case
\begin{gather*}
\bm{(-6)} \quad \wp_{666}^2  =
4\wp_{66}^3 + \wp_{56}^2 - 4\wp_{55}\wp_{66} + 4\wp_{36} - 4\wp_{45},
\\
\bm{(-7)} \quad \wp_{566}\wp_{666}  =   4\wp_{66}^2\wp_{56}
+ 2\wp_{66}^2\lambda_{6} - \wp_{55}\wp_{56} - 2\wp_{55}\lambda_{6} + 2\wp_{46}\wp_{66}
- 2\wp_{35} + \tfrac{2}{3}Q_{5556},
\\
\bm{(-8)} \quad \wp_{566}^2  =   \wp_{44} + 2\wp_{46}\lambda_{6}
+ \wp_{66}\lambda_{6}^2 + \tfrac{1}{3}Q_{5555} + \wp_{55}^2 + 4\wp_{46}\wp_{56}
+ 4\wp_{56}^2\wp_{66} + 4\lambda_{6}\wp_{56}\wp_{66},
\\
\bm{(-8)} \quad \wp_{556}\wp_{666}  =    2\wp_{56}^2\wp_{66}
+ \lambda_{6}\wp_{56}\wp_{66} - \tfrac{1}{2}\wp_{66}\lambda_{6}^2 + 2\wp_{55}\wp_{66}^2 - 2\wp_{55}^2
- \wp_{46}\wp_{56} \\
\phantom{\bm{(-8)} \quad \wp_{556}\wp_{666}  =}{}
- \lambda_{6}\wp_{46} + 4\wp_{36}\wp_{66} - 2\wp_{45}\wp_{66}
+ 2\wp_{66}\lambda_{5} + \tfrac{3}{2}\wp_{44} - \tfrac{1}{6}Q_{5555},\\
 \bm{(-9)} \quad  \wp_{556}\wp_{566}  =    2\wp_{34} + 2\wp_{55}\wp_{56}\wp_{66}
+ 2\wp_{56}\lambda_{5} + \wp_{55}\wp_{66}\lambda_{6} + 2\wp_{56}^3 + 4\wp_{36}\wp_{56} \\
\phantom{\bm{(-9)} \quad  \wp_{556}\wp_{566}  =}{}
- 2\wp_{45}\wp_{56} + \wp_{46}\wp_{55} + 2\wp_{56}^2\lambda_{6}
- 2\wp_{45}\lambda_{6} + 2\wp_{36}\lambda_{6} + 4\lambda_{4},\\
 \bm{(-9)} \quad  \wp_{555}\wp_{666}  =    6\wp_{55}\wp_{56}\wp_{66} - 8\wp_{36}\wp_{56}
- 4\wp_{35}\wp_{66} - \wp_{56}^2\lambda_{6} - 2\wp_{26} + 2\wp_{66}Q_{5556} \\
\phantom{\bm{(-9)} \quad  \wp_{555}\wp_{666}  =}{}
+ 7\wp_{45}\wp_{56} - 4\wp_{36}\lambda_{6} - 2\wp_{56}^3
+ 4\wp_{45}\lambda_{6} - 2\wp_{55}\wp_{66}\lambda_{6} + 2\wp_{46}\wp_{55},
\\
\bm{(-9)} \quad \wp_{466}\wp_{666}  =    2\wp_{55}\wp_{66}\lambda_{6}
+ 4\wp_{46}\wp_{66}^2 - 2\wp_{46}\wp_{55} + \wp_{45}\wp_{56}
- \tfrac{2}{3}\wp_{66}Q_{5556} + 2\wp_{26}, \\
\vdots
\end{gather*}
These were derived by considering the possible cubic polynomials in 2-index $\wp$-functions and using the $\sigma$-expansion to f\/ind the coef\/f\/icients.  This method fails after weight $-11$ and an alternative approach is currently being developed.  A similar problem is encountered in the (4,5)-case.

\begin{lemma}
There are a set of identities bilinear in the $2$- and $3$-index $\wp$-functions
\begin{gather}
\bm{(-6)} \quad 0=    - \tfrac{1}{2}\wp_{466} - \tfrac{1}{2}\wp_{555}
- \tfrac{1}{2}\lambda_{6}\wp_{666} - \wp_{56}\wp_{666} + \wp_{66}\wp_{566},  \label{eq:BL}
\\
\bm{(-7)} \quad 0=   - 2\wp_{366} + 2\wp_{456} - \wp_{55}\wp_{666}
- \wp_{56}\wp_{566} + 2\wp_{66}\wp_{556}, \nonumber
\\
 \bm{(-8)} \quad 0=   \tfrac{2}{3}\wp_{356} - \tfrac{2}{3}\wp_{455}
- \tfrac{2}{3}\wp_{55}\wp_{566} + \tfrac{1}{3}\wp_{56}\wp_{556} + \tfrac{1}{3}\wp_{66}\wp_{555}
- \wp_{46}\wp_{666} + \wp_{66}\wp_{466},  \nonumber
\\
 \bm{(-9)} \quad 0=   - 2\wp_{355} - \tfrac{3}{2}\lambda_{6}\wp_{466}
+ \tfrac{1}{2}\lambda_{6}\wp_{555} + \tfrac{1}{2}\lambda_{6}^2\wp_{666}
- 2\lambda_{5}\wp_{666} - \wp_{55}\wp_{556}  \nonumber \\
\phantom{\bm{(-9)} \quad 0=}{} + \wp_{56}\wp_{555} + \wp_{46}\wp_{566}
- \wp_{56}\wp_{466} - 4\wp_{36}\wp_{666} + 4\wp_{66}\wp_{366},  \nonumber
\\
\bm{(-9)} \quad 0=
2\wp_{446} - \wp_{45}\wp_{666} + \wp_{56}\wp_{466} + 2\wp_{66}\wp_{456} - 2\wp_{46}\wp_{566},  \nonumber
\\
\bm{(-10)} \ \, 0=   \tfrac{4}{3}\wp_{346}
- \wp_{266} - \tfrac{1}{3}\wp_{445} - \wp_{45}\wp_{566}
- \tfrac{1}{3}\wp_{55}\wp_{466} + \tfrac{2}{3}\wp_{56}\wp_{456} + \tfrac{2}{3}\wp_{66}\wp_{455},
 \nonumber
\\
\bm{(-10)} \ \, 0=
\tfrac{1}{2}\wp_{266} + \tfrac{1}{3}\wp_{346} + \tfrac{1}{6}\wp_{445}
- \wp_{46}\wp_{556} + \tfrac{2}{3}\wp_{55}\wp_{466}   + \tfrac{2}{3}\wp_{56}\wp_{456} - \tfrac{1}{3}\wp_{66}\wp_{455}, \nonumber
\\
 \bm{(-10)} \ \,  0=   2\wp_{346} - 2\wp_{266} - \wp_{35}\wp_{666}
+ \wp_{56}\wp_{366} + 2\wp_{66}\wp_{356} - 2\wp_{36}\wp_{566},  \nonumber\\
\vdots
 \nonumber
\end{gather}
\end{lemma}

\begin{proof}
These can be derived by cross-dif\/ferentiating the identities for 4-index $\wp$-functions.  For example, substituting equations (\ref{Bous}) and (\ref{B2}) into
\[
\frac{\partial}{\partial u_5} \wp_{6666}(\bu) - \frac{\partial}{\partial u_6} \wp_{5666}(\bu) = 0
\]
gives equation (\ref{eq:BL}).  A method to generate complete sets of such relations has been developed and will appear in a later publication.
\end{proof}

\section{Addition formula} \label{SEC_ADD}

In this section we develop the two term addition formula for the $\sigma$-function associated with the (3,7)-curve, generalising equation (\ref{eq:Intro_elliptic_add}) from the elliptic case.

\begin{theorem} \label{thm:37_add}
The $\sigma$-function associated with the cyclic {\rm (3,7)}-curve satisfies 
\[
- \frac{\sigma(\bm{u}+\bm{v})\sigma(\bm{u}-\bm{v})}{\sigma(\bm{u})^2\sigma(\bm{v})^2} = f(\bm{u},\bm{v}) + f(\bm{v},\bm{u}),
\]
where $f(\bu,\bv)$ is a finite polynomial of Abelian functions associated with $C$.
It may be written
\[
f(\bu,\bv) = \big[
P_{32} + P_{29} + P_{26} + P_{23} + P_{20} + P_{17} + P_{14} + P_{11} + P_{8} + P_{5} + P_{2}
\big](\bu,\bv),
\]
where each $P_{k}(\bu,\bv)$ is a sum of terms with weight $-k$ in the Abelian functions and weight $k-32$ in $\bm{\lambda}$-monomials.
\end{theorem}

\begin{proof}
We follow the proof of Theorem \ref{thm:37_add} for the (4,5)-case.  This time the $\sigma$-function is even instead of odd so the addition formula becomes symmetric instead of anti-symmetric.  The ratio of $\sigma$-derivatives has weight $-32$ and the polynomials have weights shifting by three since the weights of possible $\blambda$-monomials are all multiples of three.
\end{proof}

The formula has been derived explicitly using the $\sigma$-function expansion.  The same approach and Maple procedures can be used as in the (4,5)-case.  Since $f(\bu,\bv)$ may contain $\blambda$-monomials of weight $-30$, the $\sigma$-expansion should be truncated after $C_{46}$.  We f\/ind the polynomials $P_k(\bu,\bv)$ to be as given in Appendix~\ref{APP_add}.

Note that the construction of the $\sigma$-expansion, the basis (\ref{eq:37_BASIS}), the relations between Abelian functions and the addition formula are all similar to the (4,5)-results and have little in common with the lower genus trigonal cases.

\section{The cyclic trigonal curve of genus seven} \label{SEC_38}

In this section we brief\/ly present the corresponding results for the Abelian functions associated with the cyclic (3,8)-curve.  This is the non-singular algebraic curve $C$ given by
\begin{gather*}
y^3 = x^8 + \lambda_7x^7 + \lambda_6x^6 + \lambda_5x^5\lambda_4 x^4
+ \lambda_3 x^3 + \lambda_2 x^2+\lambda_1 x + \lambda_0.  
\end{gather*}
It has genus $g=7$ and is the highest genus curve to have been considered.

\subsection{Dif\/ferentials and functions}

The standard basis of holomorphic dif\/ferentials upon $C$ is
\begin{gather*}
\bm{du}  = (du_1, \dots ,du_7), \qquad \mbox{where} \quad du_i(x,y) = \frac{g_i(x,y)}{3y^2}dx, 
\end{gather*}
with
\begin{gather*}
g_1(x,y) = 1,     \qquad   g_2(x,y) = x,   \qquad   g_3(x,y) = x^2, \\
g_4(x,y) = y,   \qquad   g_5(x,y) =x^3,   \qquad   g_6(x,y) = xy,   \qquad
g_7(x,y) = x^4.
\end{gather*}
The fundamental dif\/ferential of the second kind is given by equation (\ref{eq:FundDiff}) with
\begin{gather*}
 F \big( (x,y), (z,w) \big) = 3y\lambda_{0} + \lambda_{1}yx + \lambda_{1}wz
+ \lambda_{2}x^2w + 2\lambda_{1}yz - \lambda_{4}x^4w + 2\lambda_{1}wx + 3w\lambda_{0} \nonumber  \\
 \quad  {}+ 3y^2w^2
+ \lambda_{2}yz^2 - yz^4\lambda_{4} + wx^6z^2 + 2yx^3z^5 + 2x^5z^3w
+ z^6yx^2 + 2\lambda_{2}xwz  + x^4z^3w\lambda_{7}  \nonumber \\
 \quad {} + 2\lambda_{2}xyz + 3\lambda_{3}x^2wz + 4\lambda_{4}x^3wz
+ \lambda_{5}x^4wz + 3\lambda_{3}yxz^2 + 4xyz^3\lambda_{4} + 3wx^4\lambda_{6}z^2  \nonumber \\
 \quad {}+ 2x^2yz^3\lambda_{5}
+ yz^4\lambda_{5}x + 3yz^4\lambda_{6}x^2 + yz^4\lambda_{7}x^3 + 2wx^3\lambda_{5}z^2
+ 2wx^5\lambda_{7}z^2 + 2yx^2z^5\lambda_{7}.
\end{gather*}
An explicit basis for the dif\/ferentials of the second kind associated with the cyclic (3,8)-curve was derived as follows
\begin{gather*}
\bm{dr} = (dr_1,\dots,dr_7), \qquad \mbox{where} \quad dr_j(x,y) = \frac{h_j(x,y)}{3y^2}dx,
\end{gather*}
with
\begin{gather*}
h_1(x,y) = y\big(
\lambda_2 + 13x^6 + 3\lambda_3x + 5\lambda_4x^2 + 7\lambda_5x^3 + 9\lambda_6x^4 + 11\lambda_7x^5
\big), \\
h_2(x,y) = 2xy\big( \lambda_4 + 2\lambda_5x + 3\lambda_6x^2 + 4\lambda_7x^3 + 5x^4 \big),
\qquad   h_5(x,y) = 2x^2y\big( \lambda_7 + 2x \big), \\
h_3(x,y) = y\big( \lambda_5x + 3\lambda_6x^2 + 5\lambda_7x^3 + 7x^4 - \lambda_4 \big),
\qquad h_6(x,y) = x^4\big( \lambda_7 + 2x \big),\\
h_4(x,y) = x^3\big( 5x^3 + 2\lambda_5 + 3\lambda_6x + 4\lambda_7x^2 \big),
\qquad h_7(x,y) = x^2y.
\end{gather*}
We can now proceed to def\/ine the period lattice and Abelian functions as in the general case.
This time the functions have $g=7$ variables.

\subsection{Expanding the Kleinian formula} \label{SUBSEC_38JIP}

We consider again the Kleinian formula, this time applied to the (3,8)-curve.  This is given by equation (\ref{eq:37_Klein}) but with the sum running to $g=7$.  We use the techniques of Section \ref{SEC_KF} to derive relations between the $\wp$-functions.  Expanding the formula as a series in $\xi$ gives a sequence of equations starting with the three below
\begin{gather}
0  = \rho_1 =    \wp_{77}z^4 + \wp_{57}z^3 + (\wp_{37} - w)z^2
+ (\wp_{67}w + \wp_{27})z + \wp_{47}w + \wp_{17}, \label{eq:38_pp1} \\
0  = \rho_2 =   - 2z^5 + (\wp_{67} - \wp_{777} - \lambda_{7} )z^4
+ (\wp_{56} - \wp_{577})z^3 + (\wp_{36} - \wp_{377})z^2 \nonumber \\
\phantom{0  = \rho_2 =}{}  + \big(( \wp_{66} - \wp_{677} )w - \wp_{277} + \wp_{26} \big)z
+ (\wp_{46} - \wp_{477})w - \wp_{177} + \wp_{16}, \nonumber\\ 
0  = \rho_3 =
\big( \tfrac{1}{2}\wp_{7777} - \tfrac{3}{2}\wp_{677} \big)z^4
+ \big(\tfrac{1}{2}\wp_{5777} - \tfrac{3}{2}\wp_{567}\big)z^3
+ \big( \tfrac{1}{2}\wp_{3777} - \tfrac{3}{2}\wp_{367} \big)z^2
+ \big( \tfrac{1}{2}\wp_{2777} \nonumber \\
\phantom{0  = \rho_3 =}{}
 - \tfrac{3}{2}\wp_{267} + \big(\tfrac{1}{2}\wp_{6777} - \tfrac{3}{2}\wp_{667}\big)w \big)z
- 3w^2 + \big( \tfrac{1}{2}\wp_{4777} - \tfrac{3}{2}\wp_{467} \big)w
- \tfrac{3}{2}\wp_{167} + \tfrac{1}{2}\wp_{1777}.
\nonumber
\end{gather}

The polynomials have been calculated explicitly (using Maple), up to $\rho_{8}$.  They get increasingly large in size and can be found at \cite{37web}.
We take resultants of pairs of equations, eliminating the variable $w$ by choice.  We denote the resultant of $\rho_i$ and $\rho_j$ by $\rho_{i,j}$ and f\/ind that $\rho_{1,2}$ is the simplest with 55 terms and degree seven in $z$
\begin{gather}
 \rho_{1,2} = 2z^7 + ( \lambda_{7} - 3\wp_{67} + \wp_{777})z^6 + ( \wp_{77}\wp_{677} - \wp_{77}\wp_{66} - 2\wp_{47}
- \wp_{777}\wp_{67} - \wp_{67}\lambda_{7} + \wp_{67}^2 \nonumber \\
 \phantom{\rho_{1,2} =}{} + \wp_{577} - \wp_{56})z^5 + (\wp_{377} - \wp_{77}\wp_{46} + \wp_{77}\wp_{477} - \wp_{47}\lambda_{7}
+ \wp_{57}\wp_{677} + \wp_{67}\wp_{47} + \wp_{56}\wp_{67} \nonumber \\
\phantom{\rho_{1,2} =}{} - \wp_{777}\wp_{47} - \wp_{57}\wp_{66} - \wp_{577}\wp_{67} - \wp_{36})z^4
+ ( \wp_{277} - \wp_{377}\wp_{67} - \wp_{26} - \wp_{37}\wp_{66} + \wp_{36}\wp_{67} \nonumber \\
\phantom{\rho_{1,2} =}{}
 - \wp_{577}\wp_{47} + \wp_{57}\wp_{477} + \wp_{56}\wp_{47} - \wp_{57}\wp_{46} + \wp_{37}\wp_{677})z^3 + ( \wp_{36}\wp_{47} - \wp_{377}\wp_{47}  \nonumber \\
\phantom{\rho_{1,2} =}{}
 + \wp_{27}\wp_{677} + \wp_{177} - \wp_{16} + \wp_{26}\wp_{67} + \wp_{37}\wp_{477} - \wp_{37}\wp_{46} - \wp_{27}\wp_{66}
- \wp_{277}\wp_{67})z^2  \nonumber \\
\phantom{\rho_{1,2} =}{}
+ (\wp_{17}\wp_{677} - \wp_{27}\wp_{46} + \wp_{27}\wp_{477} + \wp_{26}\wp_{47} - \wp_{177}\wp_{67} + \wp_{16}\wp_{67} - \wp_{277}\wp_{47}
- \wp_{17}\wp_{66})z  \nonumber \\
\phantom{\rho_{1,2} =}{}
- \wp_{177}\wp_{47} + \wp_{16}\wp_{47} - \wp_{17}\wp_{46} + \wp_{17}\wp_{477}. \label{eq:38_rho12}
\end{gather}
Although the genus is higher, the $\rho_{i,j}$ are not nearly as complicated as the corresponding ones in the (4,5)-case.  Also, as in the (3,7)-case, we only have to use one round of substitution to f\/ind polynomials of degree $g-1$.  We can again use these equations to solve the Jacobi inversion problem.

\begin{theorem} \label{thm:38_JIP}
Suppose we are given $\{u_1, \dots, u_7\} = \bm{u} \in J$.  Then we could solve the Jacobi inversion problem explicitly using the equations derived from the Kleinian formula.
\end{theorem}
\begin{proof}
The equation $\rho_{1,2}=0$ given by equation (\ref{eq:38_rho12}) above has degree seven in $z$.  We denote by $(z_1,\dots,z_7)$ the roots of this equation, which will be expressions in $\wp$-functions evaluated at the given $\bm{u}$.  Next consider equation (\ref{eq:38_pp1}) which is degree one in $w$.  Substitute each $z_i$ into equation (\ref{eq:38_pp1}) in turn and solve to f\/ind the corresponding $w_i$.  We have then identif\/ied the set of points
$\{(z_1, w_1), \dots, (z_7,w_7)\}$ on the curve $C$ which are the
Abel preimage of $\bm{u}$.
\end{proof}

So the manipulation of the Kleinian formula and solution of the Jacobi inversion problem seems to share a computational dif\/f\/iculty with the lower genus trigonal problems.  In fact, it is possible to easily solve the Jacobi inversion problem with this method for trigonal curves of much higher genus as demonstrated in Appendix~\ref{APP_JIP}.

\subsection[The $\sigma$-function expansion]{The $\boldsymbol{\sigma}$-function expansion}

The $\sigma$-function expansion for the cyclic (3,8)-curve may be constructed as in the (3,7)-case.

\begin{theorem} \label{thm:38_sigexp}
The function $\sigma(\bu)$ associated with the cyclic $(3,8)$-curve may be expanded about the origin as
\begin{gather*}
\sigma(\textbf{u}) = \sigma(u_1, u_2, u_3, u_4, u_5, u_6, u_7)  = SW_{3,8}(\bu) +  C_{24}(\bu) + C_{27}(\bu)
+ \dots + C_{21 + 3n}(\bu) + \cdots\!,\!\!
\end{gather*}
where $SW_{3,8}$ is the Schur--Weierstrass polynomial and each $C_k$ is a finite, odd polynomial composed of products of monomials in $\bm{u} = (u_1,u_2,\dots,u_7)$ of weight $+k$ multiplied by monomials in $\bm{\lambda}=(\lambda_7,\lambda_6,\dots,\lambda_0)$ of weight $21-k$.
\end{theorem}

\begin{proof}
Identical to the proof of Theorem~\ref{thm:37_sigexp} except the $\sigma$-function is now odd with \linebreak weight~$+21$.
\end{proof}
Following Appendix A, the Schur--Weierstrass polynomial generated by (3,8) is given by
\begin{gather*}
SW_{3,8} =  \tfrac{1}{45660160000}u_{7}^{21} + u_{3}^3  + u_{2}^2u_{7} - u_{5}^4u_{4}
+ \tfrac{1}{5}u_{7}^2u_{6}^5u_{5}u_{4}
- \tfrac{1}{10}u_{7}^5u_{6}u_{5}u_{4}^2 - \tfrac{1}{80}u_{6}^8u_{4} \nonumber \\
 \phantom{SW_{3,8} =}{}
 - \tfrac{1}{70}u_{7}^7u_{5}u_{2}
- \tfrac{2}u_{5}u_{3}u_{2}
+ \tfrac{1}{10}u_{6}^5u_{2}u_{7}
- \tfrac{1}{1120}u_{7}^9u_{2}u_{6}
+ \tfrac{1}{20}u_{6}^3u_{7}^5u_{2}
+ u_{3}^2u_{7}^2u_{4}   \nonumber \\
\phantom{SW_{3,8} =}{}
+ \tfrac{1}{1120}u_{7}^8u_{5}^2u_{4}- \tfrac{1}{8}u_{3}u_{7}^4u_{4}^2
+ \tfrac{2}{5}u_{6}^5u_{5}u_{3}
- \tfrac{1}{8}u_{6}^2u_{7}^4u_{1}
- u_{3}u_{7}u_{1}
+ u_{6}^3u_{4}u_{2}
+ \tfrac{1}{4}u_{6}^2u_{7}^4u_{5}^2u_{4}  \nonumber \\
\phantom{SW_{3,8} =}{}
+ \tfrac{1}{2}u_{5}^2u_{4}^2u_{7}^3\!
 + u_{3}u_{7}u_{5}^2u_{4}\!
- \tfrac{3}{326144000}u_{7}^{17}u_{6}^2\!
+ \tfrac{1}{1164800}u_{7}^{13}u_{6}^4\!
+ \tfrac{1}{400}u_{6}^{10}u_{7}\!
+ \tfrac{1}{280}u_{7}^8u_{6}u_{5}u_{3}  \nonumber \\
\phantom{SW_{3,8} =}{}
+ \tfrac{1}{20}u_{3}u_{7}^6u_{5}^2 - \tfrac{1}{1600}u_{6}^8u_{7}^5
- u_{7}^2u_{6}^2u_{5}^2u_{3}
- \tfrac{3}{2240}u_{7}^9u_{4}u_{3}
+ 2u_{7}u_{6}u_{5}u_{3}^2 - \tfrac{1}{25088000}u_{7}^{16}u_{4}
 \nonumber \\
 \phantom{SW_{3,8} =}{}
- \tfrac{3}{2}u_{3}u_{6}^2u_{4}^2  + \tfrac{3}{11200}u_{7}^{10}u_{6}^2u_{3}
- u_{7}^3u_{6}u_{5}u_{4}u_{3}
- \tfrac{1}{22400}u_{7}^9u_{6}^6 - \tfrac{1}{4}u_{7}u_{4}^4
- \tfrac{1}{40}u_{6}^2u_{7}^5u_{4}u_{3}
\nonumber \\
\phantom{SW_{3,8} =}{}
- \tfrac{1}{20}u_{3}u_{7}^2u_{6}^6  - \tfrac{1}{4}u_{3}u_{7}u_{6}^4u_{4}
+ \tfrac{1}{2}u_{6}^2u_{7}^3u_{3}^2
- u_{3}u_{7}^2u_{2}u_{6}
- u_{7}^2u_{5}u_{4}u_{2} + \tfrac{1}{80}u_{7}^6u_{4}^3
 \nonumber \\
 \phantom{SW_{3,8} =}{}
- 2u_{5}^2u_{2}u_{7}u_{6}
- \tfrac{1}{20}u_{5}^4u_{7}^5 - \tfrac{1}{291200}u_{7}^{13}u_{5}^2
+ \tfrac{1}{4}u_{7}^4u_{6}u_{4}u_{2}
+ u_{7}^2u_{6}u_{5}u_{1} + \tfrac{1}{4076800}u_{7}^{14}u_{3}
 \nonumber \\
 \phantom{SW_{3,8} =}{}
+ \tfrac{1}{2548000}u_{7}^{15}u_{6}u_{5}
+ \tfrac{1}{350}u_{7}^7u_{6}^5u_{5} - \tfrac{1}{40}u_{6}^4u_{7}^5u_{5}^2
- \tfrac{1}{10}u_{5}^2u_{7}u_{6}^6
+ u_{5}^4u_{7}u_{6}^2
- \tfrac{1}{1120}u_{7}^9u_{6}^2u_{5}^2
 \nonumber \\
 \phantom{SW_{3,8} =}{}
- \tfrac{1}{2}u_{5}^2u_{6}^4u_{4} + \tfrac{1}{4}u_{7}^2u_{6}^2u_{4}^3
+ u_{4}^3u_{6}u_{5} - \tfrac{1}{80}u_{6}^4u_{7}^6u_{3}
+ \tfrac{1}{2800}u_{7}^{10}u_{6}u_{5}u_{4}
+ \tfrac{1}{448}u_{7}^8u_{1}
- \tfrac{1}{4}u_{6}^4u_{1}
 \nonumber \\
 \phantom{SW_{3,8} =}{}
+ u_{5}^2u_{1} - \tfrac{1}{8}u_{6}^4u_{7}^3u_{4}^2
- \tfrac{1}{280}u_{7}^7u_{3}^2 + \tfrac{1}{80}u_{6}^6u_{7}^4u_{4}
- \tfrac{1}{4480}u_{7}^8u_{6}^4u_{4}
- \tfrac{1}{44800}u_{7}^{12}u_{6}^2u_{4}\\
\phantom{SW_{3,8} =}{}
+ \tfrac{1}{560}u_{7}^7u_{6}^2u_{4}^2
+ \tfrac{1}{22400}u_{7}^{11}u_{4}^2. \nonumber 
\end{gather*}
The other $C_{k}$ are constructed successively as in the (3,7) and (4,5)-cases.
We have calculated the expansion up to and including $C_{42}$ with the polynomials available at \cite{37web}.  The polynomials are rising in size much quicker than in the (4,5) and (3,7)-cases.  Derivation of further polynomials would be computationally dif\/f\/icult.

\subsection{Relations between the Abelian functions}

The basis for the fundamental Abelian functions has not been completed for the cyclic (3,8)-curve yet as further $\sigma$-expansion is required.  The dimension of the space is $2^g=2^7=128$ which is considerably larger than in the previous cases.  So far the 4-index $Q$-functions down to weight~$-22$ have been examined and 70 of the basis elements have been identif\/ied
\begin{gather*}  
\begin{array}{@{}ccccccccccccccc}
    & \C1         &\op& \C\wp_{11}  &\op& \C\wp_{12}  &\op& \C\wp_{13}  &\op& \C\wp_{14}
&\op& \C\wp_{15}  \\
\op & \C\wp_{16}  &\op& \C\wp_{17}  &\op& \C\wp_{22}  &\op& \C\wp_{23}  &\op& \C\wp_{24}
&\op& \C\wp_{25}  \\
\op & \C\wp_{26}  &\op& \C\wp_{27}  &\op& \C\wp_{33}  &\op& \C\wp_{34}  &\op& \C\wp_{35}
&\op& \C\wp_{36}  \\
\op & \C\wp_{37}  &\op& \C\wp_{44}  &\op& \C\wp_{45}  &\op& \C\wp_{46}  &\op& \C\wp_{47}
&\op& \C\wp_{55}  \\
\op & \C\wp_{56}  &\op& \C\wp_{57}  &\op& \C\wp_{66}  &\op& \C\wp_{67}  &\op& \C\wp_{77}
&\op& \C Q_{6667} \\
\op & \C Q_{4777} &\op& \C Q_{4667} &\op& \C Q_{5577} &\op& \C Q_{5567} &\op& \C Q_{3667}
&\op& \C Q_{4477} \\
\op & \C Q_{4467} &\op& \C Q_{4566} &\op& \C Q_{5557} &\op& \C Q_{3477} &\op& \C Q_{3567}
&\op& \C Q_{2667} \\
\op & \C Q_{3467} &\op& \C Q_{3466} &\op& \C Q_{3557} &\op& \C Q_{4447} &\op& \C Q_{4456}
&\op& \C Q_{5555} \\
\op & \C Q_{2567} &\op& \C Q_{3367} &\op& \C Q_{4555} &\op& \C Q_{1667} &\op& \C Q_{2467}
&\op& \C Q_{2566} \\
\op & \C Q_{3366} &\op& \C Q_{2377} &\op& \C Q_{2466} &\op& \C Q_{2557} &\op& \C Q_{3555}
&\op& \C Q_{4445} \\
\op & \C Q_{1477} &\op& \C Q_{1567} &\op& \C Q_{2367} &\op& \C Q_{2457} &\op& \C Q_{1377}
&\op& \C Q_{1466} \\
\op & \C Q_{1557} &\op& \C Q_{2277} &\op& \C Q_{2357} &\op& \C Q_{2555} &   &
    &             \\
\end{array}
\end{gather*}
It is likely that 6-index or even higher index $Q$-functions will be required to complete the basis.

As before we obtain equations for those $Q$-functions not in the basis as a linear combination of basis entries
\begin{gather*}
\bm{(-4)} \quad Q_{7777}  = - 3\wp_{66}, \\
\bm{(-5)} \quad Q_{6777} = 3\wp_{57}, \\
\bm{(-6)} \quad Q_{6677} = 4\wp_{47} - \wp_{56} + 2\lambda_7\wp_{67},  \\
\bm{(-7)} \quad Q_{5777} = - Q_{6667}, \\
\bm{(-8)} \quad Q_{6666} = 12\wp_{37} - 3\wp_{55} - 4Q_{4777}, \\
\bm{(-8)} \quad Q_{5677} = 4\wp_{37} - \wp_{55} + 2\lambda_7\wp_{57}, \\
\bm{(-9)} \quad Q_{5667} = 2\wp_{36} - 2\wp_{45} + 3\lambda_7\wp_{47} + 3\lambda_6\wp_{67} + \lambda_5, \\
\bm{(-9)} \quad Q_{4677} = -\wp_{45} + 2\wp_{47}\lambda_7, \\
\vdots
\end{gather*}
The relations have been calculated down to weight $-20$ and can be found at~\cite{37web}.
We can apply equation (\ref{eq:45_4iQ}) to the f\/irst set of equations in order to derive a  generalisation of equation (\ref{eq:Intro_elliptic_diff2})
\begin{gather*}
\bm{(-4)} \quad \wp_{7777} = 6\wp_{77}^2  -  3\wp_{66}, \\
\bm{(-5)} \quad \wp_{6777} = 3\wp_{57} + 6\wp_{67}\wp_{77}, \\
\bm{(-6)} \quad \wp_{6677} = 4\wp_{47} - \wp_{56} + 2\wp_{67}\lambda_{7} + 2\wp_{66}\wp_{77}
+ 4\wp_{67}^2,  \\
\vdots
\end{gather*}

\begin{remark} \label{rem:38_Bous}
As in Remark \ref{rem:37_Bous}, the f\/irst equation in this set may be dif\/ferentiated twice with respect to $u_7$ to give the Boussinesq equation for $\wp_{77}$, with $u_7$ playing the space variable and $u_6$ the time variable.

In fact, this should be the case for all the trigonal $(n,s)$ curves.  We can see this by considering the weight properties of these curves in more detail.  The trigonal curves will all have variables $\bm{u} = (u_1,\dots,u_g)$ for which $u_g$ has weight 1, $u_{g-1}$ has weight 2 and the other $u_j$ have weight greater than 3.  (This is clear from the derivation of the weights discussed in detail in~\cite{MEe09}.)

The function $Q_{gggg}$ will be an Abelian function associated to the curve with poles of order two and weight $-4$.  Since it is the only such Q-function it should be expressible as a linear combination of the 2-index $\wp$-functions at this weight.  Given the discussion of weights above we see that the only such function is $\wp_{g-1,g-1}$.  Hence we should f\/ind an equation $Q_{gggg} = c\wp_{g-1,g-1}$ for a suitable constant $c$.  Upon substituting for the Q-function and dif\/ferentiating twice we will be left with a Boussinesq type equation.
\end{remark}

An addition formula for the $(3,8)$-curve has not been constructed as it requires the basis to be completed and more of the $\sigma$-expansion to be derived.  Further progress here is limited by the computational time and memory required for calculations.  However, it is clear that the higher genus will mean the results are further removed from the lower genus cases.

\section{Conclusions}

The dif\/f\/iculty and complexity of calculating the $\sigma$-expansion and deriving the relations between $\wp$-functions is dictated by the genus of the curve considered.  However, the expansion of the Kleinian formula provides an exception to this rule, and we f\/ind that some results derived
from this such as the solution to the Jacobi inversion problem and the connection with the Boussinesq equation are similar for trigonal curves of arbitrary genus.

\appendix

\section{Constructing Schur--Weierstrass polynomials} \label{APP_SW}

In this Appendix we summarise the algorithmic procedure that can be used to calculate Schur--Weierstrass polynomials.  For more information on these polynomials and their role in the theory see \cite{bel99} and \cite{N08}.  To construct the Schur--Weierstrass polynomial associated to $(n,s)$:
\begin{enumerate}\itemsep=0pt
\item Calculate the Weierstrass gap sequence generated by $(n,s)$.  These are the natural numbers not representable in the form $an+bs$ where $a,b\in\N$.  There will be $g$ such numbers, $W_{n,s} = \{w_1,\dots,w_g\}$ where $g$ corresponds with the genus of the associated $(n,s)$-curve.
\item Calculate the associated Weierstrass partition, $\pi_{n,s}$ as the partition in which each entry is given by $\pi_k = w_{g-k+1}+k-g$ for $k-1,\dots,g$.
\item Write the elementary symmetric polynomials using their representation in elementary Newton polynomials, given by
\[
e_k = \frac{1}{k!} \left|
\begin{array}{lllllll}
p_1     & 1       & 0       & \dots  & 0      & 0      & 0     \\
p_2     & p_1     & 2       & \dots  & 0      & 0      & 0     \\
\vdots  & \vdots  & \vdots  & \ddots & \vdots & \vdots & \vdots \\
p_{k-1} & p_{k-2} & p_{k-2} & \dots  & p_2    & p_1    & k-1   \\
p_k     & p_{k-1} & p_{k-2} & \dots  & p_3    & p_2    & p_1
\end{array}
\right|.
\]
\item Denote the Schur polynomial def\/ined by the Weierstrass partition as $S_{n,s}$ and calculate it as
$S_{n,s} = \det \big( (e_{\pi_i-i-j})_{1\leq i,j\leq g} \big)$, where the $e_k$ are as given in the previous step.
\item Denote the corresponding Schur--Weierstrass polynomial by $SW_{n,s}$ and derive it by making the change of variables $p_{w_i}=w_i u_{g+1-i}$, $i=1,\dots,g$ into $S_{n,s}$.
\end{enumerate}

\section{Addition formula associated to the cyclic (3,7)-curve} \label{APP_add}

The polynomial $f(\bu,\bv)$ in Theorem \ref{thm:37_add} is given by
\[
f(\bu,\bv) = \big[
P_{32} + P_{29} + P_{26} + P_{23} + P_{20} + P_{17} + P_{14} + P_{11} + P_{8} + P_{5} + P_{2}
\big](\bu,\bv),
\]
where the polynomials $P_k(\bu,\bv)$ are given as follows
\begin{gather*}
P_{2}(\bu,\bv) =
- \tfrac{4}{63}\wp_{66}(\bv)\lambda_{6}\lambda_{5}\lambda_{4}\lambda_{3}
- \tfrac{3}{2}\wp_{66}(\bv)\lambda_{4}\lambda_{0}
- \tfrac{1}{3}\wp_{66}(\bu)\lambda_{5}^2\lambda_{1}
- \tfrac{9}{4}\wp_{66}(\bv)\lambda_{6}^3\lambda_{0} \\
\quad{}
+ \tfrac{4}{3}\wp_{66}(\bu)\lambda_{3}\lambda_{1}
- 3\wp_{66}(\bv)\lambda_{5}\lambda_{4}\lambda_{2}
+ \tfrac{3}{4}\wp_{66}(\bv)\lambda_{6}^2\lambda_{4}\lambda_{2}
+ \tfrac{1}{3}\wp_{66}(\bu)\lambda_{2}^2
+ 9\wp_{66}(\bv)\lambda_{6}\lambda_{5}\lambda_{0}  \\
\quad{}
+ \tfrac{1}{12}\wp_{66}(\bu)\lambda_{6}^2\lambda_{5}\lambda_{1}
+ \tfrac{7}{6}\wp_{66}(\bv)\lambda_{6}\lambda_{4}\lambda_{1}
- \tfrac{41}{21}\wp_{66}(\bv)\lambda_{4}^2\lambda_{3}
- \tfrac{25}{42}\wp_{66}(\bu)\lambda_{6}\lambda_{4}^3 \\
\quad{}
+ \tfrac{5}{21}\wp_{66}(\bv)\lambda_{5}^2\lambda_{4}^2
+ \tfrac{1}{63}\wp_{66}(\bu)\lambda_{6}\lambda_{5}^3\lambda_{4}
- \tfrac{1}{252}\wp_{66}(\bu)\lambda_{6}^3\lambda_{5}^2\lambda_{4}
- \tfrac{25}{252}\wp_{66}(\bu)\lambda_{6}^2\lambda_{5}\lambda_{4}^2,\\[0.5ex]
P_{5}(\bu,\bv) =   \big[ \tfrac{25}{42}\wp_{56}(\bu)\wp_{66}(\bv)
- \tfrac{3}{2}\wp_{46}(\bv) \big] \lambda_{6}\lambda_{4}\lambda_{2}
+ \big[ \tfrac{20}{21}\wp_{66}(\bu)\wp_{56}(\bv) - \wp_{46}(\bu) \big] \lambda_{3}\lambda_{2} \\
\quad{} + \tfrac{5}{84}\wp_{56}(\bu)\wp_{66}(\bv)\lambda_{6}^2\lambda_{5}\lambda_{2}
+ \big[ \tfrac{15}{2}\wp_{56}(\bu)\wp_{66}(\bv) - 3\wp_{46}(\bu) \big]\lambda_{5}\lambda_{0}
+ \tfrac{13}{6}\wp_{46}(\bv)\lambda_{6}\lambda_{5}\lambda_{1} \\
\quad{}
+ \big[ \tfrac{5}{3}\wp_{56}(\bu)\wp_{66}(\bv) + \tfrac{11}{3}\wp_{46}(\bv) \big]\lambda_{4}\lambda_{1}
+ \big[ \tfrac{9}{2}\wp_{46}(\bv) - \tfrac{15}{8}\wp_{66}(\bu)\wp_{56}(\bv) \big]\lambda_{6}^2\lambda_{0}   \\
\quad{}
- \tfrac{2}{63}\wp_{46}(\bu)\lambda_{6}^2\lambda_{5}^2\lambda_{4}
- \tfrac{11}{42}\wp_{46}(\bv)\lambda_{6}\lambda_{5}\lambda_{4}^2
- \big[ \tfrac{1}{2}\wp_{46}(\bv) + \tfrac{5}{21}\wp_{56}(\bu)\wp_{66}(\bv) \big]\lambda_{5}^2\lambda_{2} \\
\quad{}
- \tfrac{25}{14}\wp_{46}(\bv)\lambda_{4}^3 - \tfrac{5}{2}\wp_{46}(\bv)\lambda_{5}\lambda_{4}\lambda_{3},\\[0.5ex]
P_{8}(\bu,\bv) =
- \big[ \tfrac{221}{126}\wp_{45}(\bv)\wp_{66}(\bu)
+ \tfrac{8}{63}\wp_{36}(\bu)\wp_{66}(\bv) \big] \lambda_{6}\lambda_{4}\lambda_{3}
+ \tfrac{5}{7}\wp_{45}(\bu)\wp_{66}(\bv)\lambda_{5}^2\lambda_{3} \\
\quad{}
+ \big[ \tfrac{1}{4}\wp_{45}(\bv)\wp_{66}(\bu)
+ \tfrac{1}{4}Q_{5555}(\bu) + \tfrac{3}{4}\wp_{44}(\bu)
+ \tfrac{103}{42}\wp_{46}(\bu)\wp_{56}(\bv) \big]\lambda_{4}\lambda_{2}
+ \big[ \tfrac{5}{12}\wp_{44}(\bu) \\
\quad{}   + 3\wp_{36}(\bu)\wp_{66}(\bv)
+ \tfrac{1}{36}Q_{5555}(\bu) - 4\wp_{45}(\bu)\wp_{66}(\bv)
+ \tfrac{7}{3}\wp_{56}(\bv)\wp_{46}(\bu) \big]\lambda_{5}\lambda_{1} \\
\quad{}
+ \big[ \wp_{45}(\bu)\wp_{66}(\bv) - \tfrac{3}{4}\wp_{36}(\bu)\wp_{66}(\bv) \big]\lambda_{6}^2\lambda_{1} + \big[ \tfrac{2}{63}\wp_{36}(\bu)\wp_{66}(\bv) - \tfrac{1}{126}\wp_{45}(\bu)\wp_{66}(\bv)
\\
\quad{}
- \tfrac{1}{756}Q_{5555}(\bu)
- \tfrac{1}{84}\wp_{44}(\bv)\big]\lambda_{6}\lambda_{5}^2\lambda_{4}
- \big[ \tfrac{5}{28}\wp_{44}(\bv) + \tfrac{5}{252}Q_{5555}(\bv) \big]\lambda_{5}\lambda_{4}^2
+ \big[ - \tfrac{3}{4}Q_{5555}(\bv)\\
\quad{}   + \tfrac{15}{4}\wp_{56}(\bu)\wp_{46}(\bv)  \big]\lambda_{6}\lambda_{0}
- \tfrac{5}{28}\wp_{66}(\bu)\wp_{45}(\bv)\lambda_{6}^2\lambda_{5}\lambda_{3}
- \big[ 5\wp_{55}(\bu)\wp_{55}(\bv) + \tfrac{33}{4}\wp_{44}(\bu) \big]\lambda_{6}\lambda_{0}  \\
\quad{}   + \big[ \tfrac{5}{252}\wp_{66}(\bv)\wp_{45}(\bu)
- \tfrac{5}{63}\wp_{66}(\bu)\wp_{36}(\bv) \big]\lambda_{6}^2\lambda_{4}^2
+ \tfrac{10}{21}\wp_{46}(\bu)\wp_{56}(\bv)\lambda_{6}\lambda_{5}\lambda_{2} \\
\quad{}
+ \big[ \tfrac{1}{504}\wp_{66}(\bu)\wp_{45}(\bv)
- \tfrac{1}{126}\wp_{36}(\bu)\wp_{66}(\bv) \big] \lambda_{6}^3\lambda_{5}\lambda_{4}
- \tfrac{20}{7}\wp_{45}(\bu)\wp_{66}(\bv)\lambda_{3}^2,
\\[0.5ex]
P_{11}(\bu,\bv) =
- \tfrac{1}{252}\big[ \wp_{66}(\bu)\wp_{34}(\bv)
+ \wp_{26}(\bu)\wp_{66}(\bv)\big]\lambda_{5}^2\lambda_{6}^3
+ \big( \big[\tfrac{1}{63}Q_{3466}(\bu) + \tfrac{1}{63}Q_{4456}(\bu)  \\
\quad{}   - \tfrac{4}{63}\wp_{36}(\bu)\wp_{46}(\bv)
- \tfrac{41}{504}\wp_{26}(\bu)\wp_{66}(\bv) + \tfrac{1}{63}\wp_{46}(\bu)\wp_{45}(\bv)
- \tfrac{37}{252}\wp_{66}(\bu)\wp_{34}(\bv)\big]\lambda_{4}\lambda_{5}\\
\quad{}
+ \big[\tfrac{1}{2}\wp_{34}(\bu)\wp_{66}(\bv)
- \tfrac{1}{8}\wp_{66}(\bu)\wp_{26}(\bv)\big]\lambda_{2}  \big)\lambda_{6}^2
+ \big( \big[ \tfrac{5}{21}Q_{3466}(\bv) - \tfrac{5}{12}\wp_{66}(\bu)\wp_{26}(\bv)   \\
\quad{}   - \tfrac{3}{28}\wp_{45}(\bu)\wp_{46}(\bv) + \tfrac{5}{21}Q_{4456}(\bv)
- \tfrac{15}{14}\wp_{66}(\bu)\wp_{34}(\bv)
+ \tfrac{3}{7}\wp_{46}(\bu)\wp_{36}(\bv)\big]\lambda_{4}^2  \\
\quad{}   + \big[\tfrac{1}{63}\wp_{66}(\bu)\wp_{34}(\bv)
+ \tfrac{1}{63}\wp_{26}(\bu)\wp_{66}(\bv)\big]\lambda_{5}^3
+ \big[\tfrac{1}{12}\wp_{26}(\bu)\wp_{66}(\bv) - \wp_{45}(\bu)\wp_{46}(\bv) \\
\quad{}   - \tfrac{5}{6}Q_{3466}(\bu) + 4\wp_{36}(\bu)\wp_{46}(\bv)
- 4\wp_{35}(\bu)\wp_{55}(\bv) - \tfrac{2}{3}\wp_{34}(\bu)\wp_{66}(\bv)
- \tfrac{5}{6}Q_{4456}(\bu)\big]\lambda_{1} \\
\quad{}   - \big[ \tfrac{4}{63}\wp_{26}(\bu)\wp_{66}(\bv)
+ \tfrac{4}{63}\wp_{34}(\bu)\wp_{66}(\bv)
+ \tfrac{10}{7}\wp_{45}(\bu)\wp_{46}(\bv)\big]\lambda_{3}\lambda_{5}\big)\lambda_{6}
+ \big[ \tfrac{3}{2}Q_{4456}(\bu) \\
\quad{}   - \tfrac{1}{2}\wp_{34}(\bu)\wp_{66}(\bv)
+ 5\wp_{46}(\bu)\wp_{45}(\bv) - \tfrac{5}{8}Q_{5555}(\bu)\wp_{56}(\bv)
- \tfrac{47}{8}\wp_{44}(\bu)\wp_{56}(\bv) \\
\quad{}
+ \tfrac{5}{3}Q_{5556}(\bu)\wp_{55}(\bv) - 4\wp_{26}(\bu)\wp_{66}(\bv)
- \tfrac{5}{2}\wp_{35}(\bu)\wp_{55}(\bv)
+ \tfrac{17}{2}\wp_{46}(\bu)\wp_{36}(\bv) \\
\quad{}   + \tfrac{1}{2}Q_{3466}(\bu) \big]\lambda_{0}
+ \big[\tfrac{1}{6}\wp_{26}(\bu)\wp_{66}(\bv)
+ \tfrac{3}{7}\wp_{34}(\bu)\wp_{66}(\bv)\big]\lambda_{4}\lambda_{5}^2
+ \big[ \tfrac{1}{6}\wp_{36}(\bu)\wp_{46}(\bv)    \\
\quad{}   + \tfrac{5}{252}\wp_{56}(\bu)Q_{5555}(\bv) + \tfrac{7}{12}\wp_{26}(\bu)\wp_{66}(\bv)
- \tfrac{5}{6}\wp_{45}(\bu)\wp_{46}(\bv)
+ \tfrac{5}{28}\wp_{56}(\bu)\wp_{44}(\bv)  \\
\quad{}    - \tfrac{5}{3}\wp_{34}(\bu)\wp_{66}(\bv)
+ \tfrac{1}{6}Q_{4456}(\bv)
+ \tfrac{1}{6}Q_{3466}(\bv) \big]\lambda_{2}\lambda_{5}
+ \big[ \tfrac{1}{3}\wp_{66}(\bu)\wp_{26}(\bv) + \tfrac{1}{2}Q_{4456}(\bv) \\
\quad{}   - \wp_{36}(\bu)\wp_{46}(\bv)
- \tfrac{19}{7}\wp_{34}(\bu)\wp_{66}(\bv)
+ \tfrac{1}{2}Q_{3466}(\bu)
- \tfrac{199}{28}\wp_{46}(\bu)\wp_{45}(\bv)\big]\lambda_{3}\lambda_{4},
 \\[0.5ex]
P_{14}(\bu,\bv) =
\big[\tfrac{1}{84}\wp_{24}(\bu)\wp_{66}(\bv)
- \tfrac{1}{84}\wp_{16}(\bu)\wp_{66}(\bv)\big]\lambda_{5}\lambda_{6}^4
+ \big[ \tfrac{5}{42}\wp_{24}(\bu)\wp_{66}(\bv)  \\
\quad{}   - \tfrac{5}{42}\wp_{16}(\bv)\wp_{66}(\bu) \big]\lambda_{4}\lambda_{6}^3
+ \big(  \big[\tfrac{95}{168}\wp_{24}(\bu)\wp_{66}(\bv)
+ \tfrac{47}{84}\wp_{16}(\bu)\wp_{66}(\bv)\big]\lambda_{3} \\
\quad{}    - \big[ \tfrac{1}{21}\wp_{16}(\bv)\wp_{66}(\bu)
+ \tfrac{2}{63}\wp_{34}(\bv)\wp_{46}(\bu) + \tfrac{5}{84}\wp_{24}(\bv)\wp_{66}(\bu)
+ \tfrac{2}{63}\wp_{46}(\bv)\wp_{26}(\bu)\big]\lambda_{5}^2  \big)\lambda_{6}^2 \\
\quad{}
+ \big(  \big[\tfrac{9}{4}\wp_{25}(\bu)\wp_{55}(\bv) - \tfrac{5}{12}Q_{4466}(\bu)\wp_{55}(\bv)
- \tfrac{5}{21}\wp_{56}(\bu)Q_{4456}(\bv) - \tfrac{5}{21}\wp_{56}(\bu)Q_{3466}(\bv) \\
\quad{}   - \tfrac{1}{2}\wp_{24}(\bv)\wp_{66}(\bu)
+ \tfrac{3}{2}\wp_{16}(\bu)\wp_{66}(\bv) - \tfrac{1}{2}\wp_{46}(\bu)\wp_{26}(\bv)
- \wp_{55}(\bu)\wp_{33}(\bv)  \\
\quad{}   - \wp_{34}(\bv)\wp_{46}(\bu)\big]\lambda_{2}
+ \big[ \tfrac{1}{1512}\wp_{45}(\bu)Q_{5555}(\bv)
- \tfrac{5}{42}\wp_{46}(\bv)\wp_{26}(\bu) - \tfrac{9}{14}\wp_{34}(\bv)\wp_{46}(\bu)  \\
\quad{}   - \tfrac{1}{21}Q_{3446}(\bv)
- \tfrac{1}{42}\wp_{36}(\bu)\wp_{44}(\bv) + \tfrac{1}{27}Q_{4445}(\bu)
- \tfrac{1}{378}\wp_{36}(\bu)Q_{5555}(\bv) + \tfrac{1}{126}Q_{3355}(\bv) \\
\quad{}   - \tfrac{5}{42}\wp_{24}(\bv)\wp_{66}(\bu)
- \tfrac{20}{21}\wp_{16}(\bv)\wp_{66}(\bu)
+ \tfrac{1}{168}\wp_{45}(\bu)\wp_{44}(\bv)\big]\lambda_{4}\lambda_{5}  \big)\lambda_{6}  \\
\quad{}   + \big[  \tfrac{8}{21}\wp_{16}(\bu)\wp_{66}(\bv)
+ \tfrac{1}{21}\wp_{24}(\bu)\wp_{66}(\bv)  \big]\lambda_{5}^3
+ \big[ \tfrac{2}{3}\wp_{16}(\bv)\wp_{66}(\bu)
- \tfrac{4}{9}Q_{4445}(\bu)  \\
\quad{}   + \tfrac{2}{7}Q_{3446}(\bu) - \tfrac{107}{42}\wp_{34}(\bv)\wp_{46}(\bu)
- \tfrac{4}{3}\wp_{24}(\bv)\wp_{66}(\bu)
- \tfrac{25}{4}\wp_{46}(\bv)\wp_{26}(\bu)  \\
\quad{}   + \tfrac{5}{42}Q_{3355}(\bv) \big]\lambda_{4}^2
+ \big[ \tfrac{1}{4}\wp_{46}(\bv)\wp_{26}(\bu)
- \tfrac{95}{21}\wp_{16}(\bu)\wp_{66}(\bv) - \tfrac{71}{42}\wp_{24}(\bu)\wp_{66}(\bv) \\
\quad{}   - \tfrac{5}{2}\wp_{46}(\bv)\wp_{34}(\bu)
- \tfrac{5}{84}Q_{5555}(\bu)\wp_{45}(\bv)
- \tfrac{15}{28}\wp_{45}(\bu)\wp_{44}(\bv) \big]\lambda_{3}\lambda_{5}
+ \big[ \tfrac{5}{9}Q_{4445}(\bv) \\
\quad{}   - \tfrac{1}{6}Q_{3355}(\bu)
+ \tfrac{5}{4}\wp_{25}(\bv)\wp_{55}(\bu)
- \tfrac{1}{4}Q_{4466}(\bu)\wp_{55}(\bv)
- 3\wp_{35}(\bu)\wp_{35}(\bv) - \tfrac{2}{3}Q_{3446}(\bu) \\
\quad{}   - \tfrac{1}{3}\wp_{16}(\bu)\wp_{66}(\bv) - \tfrac{8}{3}\wp_{46}(\bv)\wp_{34}(\bu)
- \tfrac{11}{12}\wp_{36}(\bv)\wp_{44}(\bu)
- \tfrac{1}{4}\wp_{36}(\bu)Q_{5555}(\bv)  \\
\quad{}
- \tfrac{5}{6}Q_{3466}(\bu)\wp_{56}(\bv)
+ \tfrac{1}{12}Q_{4455}(\bu)\wp_{66}(\bv)
- \tfrac{2}{3}Q_{4456}(\bu)\wp_{56}(\bv)
- \wp_{46}(\bv)\wp_{26}(\bu)   \\
\quad{}   + \tfrac{4}{3}Q_{5556}(\bu)\wp_{35}(\bv)
+ \tfrac{11}{3}\wp_{44}(\bu)\wp_{45}(\bv)
+ \tfrac{11}{4}\wp_{24}(\bu)\wp_{66}(\bv)
+ \tfrac{1}{3}Q_{5555}(\bu)\wp_{45}(\bv)  \big]\lambda_{1},
\\[0.5ex]
P_{17}(\bu,\bv) =
\big[ \tfrac{1}{504}Q_{3346}(\bu)\wp_{66}(\bv) - \tfrac{2}{21}\wp_{16}(\bu)\wp_{46}(\bv)
- \tfrac{19}{252}\wp_{66}(\bv)\wp_{14}(\bu) \\
\quad{}   + \tfrac{2}{21}\wp_{24}(\bv)\wp_{46}(\bu)  \big]\lambda_{5}\lambda_{6}^3
+ \big[\tfrac{5}{252}\wp_{66}(\bu)Q_{3346}(\bv) - \tfrac{1}{126}Q_{3466}(\bu)\wp_{45}(\bv)
+ \wp_{15}(\bu)\wp_{55}(\bv)   \\
\quad{}
+ \tfrac{47}{63}\wp_{66}(\bv)\wp_{14}(\bu) - \tfrac{9}{14}\wp_{24}(\bu)\wp_{46}(\bv)
+ \tfrac{9}{14}\wp_{16}(\bu)\wp_{46}(\bv)  + \tfrac{2}{63}Q_{4456}(\bu)\wp_{36}(\bv)  \\
\quad{}
- \tfrac{1}{126}Q_{4456}(\bu)\wp_{45}(\bv)
+ \tfrac{2}{63}Q_{3466}(\bu)\wp_{36}(\bv)\big]\lambda_{4}\lambda_{6}^2
+ \big(  \big[ \tfrac{19}{63}\wp_{66}(\bv)\wp_{14}(\bu) \\
\quad{}   - \tfrac{1}{126}Q_{3346}(\bu)\wp_{66}(\bv)
- \tfrac{1}{756}\wp_{34}(\bv)Q_{5555}(\bu)
- \tfrac{16}{21}\wp_{16}(\bu)\wp_{46}(\bv)
- \tfrac{1}{84}\wp_{34}(\bu)\wp_{44}(\bv)    \\
\quad{}   - \tfrac{2}{21}\wp_{24}(\bu)\wp_{46}(\bv)
- \tfrac{1}{84}\wp_{44}(\bu)\wp_{26}(\bv)
- \tfrac{1}{756}\wp_{26}(\bu)Q_{5555}(\bv)\big]\lambda_{5}^2
+ \big[ \tfrac{1}{3}Q_{4446}(\bu)\wp_{55}(\bv)      \\
\quad{}   + \tfrac{2}{63}Q_{3346}(\bu)\wp_{66}(\bv)
- 6\wp_{16}(\bu)\wp_{46}(\bv)
- \tfrac{76}{63}\wp_{66}(\bv)\wp_{14}(\bu)
+ \tfrac{5}{7}\wp_{45}(\bu)Q_{3466}(\bv) \\
\quad{}   + \tfrac{3}{2}\wp_{23}(\bu)\wp_{55}(\bv)
+ \tfrac{5}{7}Q_{4456}(\bu)\wp_{45}(\bv)
+ \tfrac{3}{4}\wp_{24}(\bv)\wp_{46}(\bu)
+ 2\wp_{15}(\bu)\wp_{55}(\bv) \big]\lambda_{3}  \big)\lambda_{6} \\
\quad{}   + \big[\tfrac{1}{3}Q_{2446}(\bu)
- \tfrac{1}{8}\wp_{44}(\bu)\wp_{26}(\bv)
- \tfrac{1}{28}\wp_{34}(\bv)Q_{5555}(\bu)
- \tfrac{1}{72}\wp_{26}(\bu)Q_{5555}(\bv)   \\
\quad{}   - \tfrac{28}{3}\wp_{66}(\bu)\wp_{14}(\bv)
- \tfrac{9}{28}\wp_{34}(\bu)\wp_{44}(\bv)
- \tfrac{55}{7}\wp_{16}(\bu)\wp_{46}(\bv)
- \tfrac{11}{21}\wp_{24}(\bu)\wp_{46}(\bv)  \\
\quad{}   - \tfrac{1}{9}Q_{1555}(\bv) \big]\lambda_{4}\lambda_{5}
+ \big[ \tfrac{1}{6}Q_{3466}(\bu)\wp_{36}(\bv)
+ \tfrac{1}{3}\wp_{33}(\bu)Q_{5556}(\bv)
+ \tfrac{1}{6}Q_{3466}(\bu)\wp_{45}(\bv)  \\
\quad{}   - 2\wp_{15}(\bu)\wp_{55}(\bv)
+ \tfrac{1}{6}\wp_{34}(\bu)Q_{5555}(\bv)
+ \tfrac{9}{4}\wp_{25}(\bu)\wp_{35}(\bv)
- \tfrac{5}{12}Q_{4455}(\bu)\wp_{46}(\bv) \\
\quad{}   - \tfrac{3}{4}Q_{4466}(\bv)\wp_{35}(\bu)
- \wp_{33}(\bu)\wp_{35}(\bv)
- \tfrac{5}{2}\wp_{66}(\bv)\wp_{14}(\bu)
- \tfrac{1}{24}\wp_{26}(\bu)Q_{5555}(\bv)  \\
\quad{}    + \tfrac{1}{3}Q_{4446}(\bu)\wp_{55}(\bv)
- \tfrac{1}{8}\wp_{44}(\bv)\wp_{26}(\bu)
+ \tfrac{7}{2}\wp_{16}(\bu)\wp_{46}(\bv)
+ \tfrac{5}{18}Q_{4445}(\bu)\wp_{56}(\bv) \\
\quad{}    + \tfrac{1}{4}\wp_{24}(\bu)\wp_{46}(\bv)
- \tfrac{1}{6}Q_{4456}(\bu)\wp_{36}(\bv)
- \tfrac{2}{7}Q_{3446}(\bu)\wp_{56}(\bv)
+ \tfrac{1}{2}\wp_{34}(\bu)\wp_{44}(\bv) \\
\quad{}   + \tfrac{5}{36}Q_{4466}(\bv)Q_{5556}(\bu)
- \tfrac{1}{6}Q_{3445}(\bu)\wp_{66}(\bv)
- \tfrac{5}{42}Q_{3355}(\bu)\wp_{56}(\bv)  + \tfrac{1}{3}Q_{2446}(\bu) \\
\quad{}   - \tfrac{1}{12}Q_{3346}(\bu)\wp_{66}(\bv)
+ \tfrac{2}{3}Q_{4456}(\bu)\wp_{45}(\bv)
- \tfrac{3}{4}\wp_{25}(\bu)Q_{5556}(\bv) \big]\lambda_{2},
 \\[0.5ex]
P_{20}(\bu,\bv) =
\big[ \tfrac{1}{84}\wp_{66}(\bu)Q_{3344}(\bv)
+ \tfrac{1}{28}\wp_{66}(\bu)Q_{2445}(\bv)
+ \tfrac{1}{63}Q_{4456}(\bu)\wp_{34}(\bv) \\
\quad{}   - \tfrac{38}{63}\wp_{46}(\bu)\wp_{14}(\bv)
+ \tfrac{1}{63}\wp_{46}(\bu)Q_{3346}(\bv)
+ \tfrac{1}{63}Q_{3466}(\bu)\wp_{34}(\bv)  + \tfrac{1}{252}\wp_{24}(\bu)Q_{5555}(\bv)  \\
\quad{}   + \tfrac{1}{28}\wp_{24}(\bu)\wp_{44}(\bv)
+ \tfrac{1}{63}Q_{3466}(\bu)\wp_{26}(\bv)
+ \tfrac{1}{84}\wp_{66}(\bu)Q_{2346}(\bv) - \tfrac{1}{252}\wp_{16}(\bv)Q_{5555}(\bu) \\
\quad{}   - \tfrac{1}{28}\wp_{16}(\bu)\wp_{44}(\bv)
+ \tfrac{1}{63}Q_{4456}(\bu)\wp_{26}(\bv) \big]\lambda_{5}\lambda_{6}^2
+ \big[ \tfrac{1}{6}Q_{4456}(\bu)\wp_{26}(\bv)
- \tfrac{1}{3}Q_{1446}(\bv)   \\
\quad{}   + \tfrac{3}{7}Q_{3466}(\bu)\wp_{34}(\bv)
+ \tfrac{5}{14}\wp_{66}(\bu)Q_{2445}(\bv)
+ \tfrac{1}{42}Q_{3446}(\bv)\wp_{45}(\bu) + \wp_{15}(\bv)\wp_{35}(\bu) \\
\quad{}   + \tfrac{2}{27}Q_{4445}(\bu)\wp_{36}(\bv)
- \tfrac{2}{21}Q_{3446}(\bu)\wp_{36}(\bv)
- \tfrac{3}{28}Q_{3346}(\bu)\wp_{46}(\bv) - \tfrac{139}{14}\wp_{14}(\bu)\wp_{46}(\bv)  \\
\quad{}   - \tfrac{1}{3}\wp_{15}(\bu)Q_{5556}(\bv)
+ \tfrac{5}{42}Q_{2346}(\bu)\wp_{66}(\bv)
+ \tfrac{1}{6}\wp_{26}(\bu)Q_{3466}(\bv)
- \tfrac{1}{3}Q_{1556}(\bv)\wp_{55}(\bu)  \\
\quad{}   - \tfrac{1}{252}Q_{3355}(\bu)\wp_{45}(\bv)
+ \tfrac{1}{63}Q_{3355}(\bu)\wp_{36}(\bv)
+ \tfrac{4}{3}\wp_{13}(\bu)\wp_{55}(\bv) + \tfrac{3}{7}Q_{4456}(\bv)\wp_{34}(\bu)  \\
\quad{}   + \wp_{22}(\bu)\wp_{55}(\bv)
- \tfrac{1}{54}Q_{4445}(\bu)\wp_{45}(\bv)
+ \tfrac{5}{42}\wp_{66}(\bu)Q_{3344}(\bv)  \big]\lambda_{4}\lambda_{6}
- \big[ \tfrac{11}{3}\wp_{14}(\bu)\wp_{46}(\bv) \\
\quad{}   + \tfrac{1}{21}\wp_{66}(\bv)Q_{3344}(\bu)
+ \tfrac{1}{252}\wp_{24}(\bv)Q_{5555}(\bu)
+ \tfrac{2}{63}\wp_{16}(\bu)Q_{5555}(\bv)
+ \tfrac{1}{21}Q_{2346}(\bu)\wp_{66}(\bv) \\
\quad{}   + \tfrac{1}{28}\wp_{24}(\bu)\wp_{44}(\bv)
+ \tfrac{1}{7}Q_{2445}(\bu)\wp_{66}(\bv)
+ \tfrac{2}{7}\wp_{16}(\bu)\wp_{44}(\bv)  \big]\lambda_{5}^2 \\
\quad{}   + \big[ \tfrac{4}{21}\wp_{66}(\bu)Q_{2346}(\bv)
- \tfrac{3}{4}\wp_{25}(\bu)\wp_{25}(\bv)
+ \tfrac{3}{2}\wp_{23}(\bu)\wp_{35}(\bv)
+ \tfrac{3}{8}\wp_{24}(\bu)\wp_{44}(\bv)    \\
\quad{}   + \tfrac{5}{4}\wp_{16}(\bu)\wp_{44}(\bv)
+ \tfrac{1}{8}\wp_{24}(\bu)Q_{5555}(\bv)
+ \tfrac{1}{2}Q_{3466}(\bv)\wp_{34}(\bu)
- \tfrac{1}{2}Q_{3466}(\bu)\wp_{26}(\bv)  \\
\quad{}   - \tfrac{1}{2}Q_{5556}(\bu)\wp_{23}(\bv)
+ \tfrac{6}{7}Q_{3446}(\bu)\wp_{45}(\bv)
+ Q_{4466}(\bu)\wp_{25}(\bv)  + \tfrac{5}{14}\wp_{45}(\bu)Q_{3355}(\bv) \\
\quad{}    + \tfrac{4}{21}\wp_{66}(\bv)Q_{3344}(\bu)
- \tfrac{1}{6}\wp_{35}(\bv)Q_{4446}(\bu) + 3\wp_{15}(\bu)\wp_{35}(\bv)
- \tfrac{1}{9}Q_{5556}(\bu)Q_{4446}(\bv)  \\
\quad{}   + \tfrac{1}{2}Q_{4456}(\bu)\wp_{34}(\bv)
- \tfrac{1}{12}Q_{4466}(\bu)Q_{4466}(\bv)
- \tfrac{13}{2}\wp_{46}(\bu)\wp_{14}(\bv)
- \tfrac{2}{3}\wp_{15}(\bu)Q_{5556}(\bv)  \\
\quad{}   - \tfrac{5}{6}\wp_{45}(\bu)Q_{4445}(\bv)
+ \tfrac{1}{4}\wp_{16}(\bu)Q_{5555}(\bv)
+ \tfrac{1}{4}\wp_{46}(\bu)Q_{3346}(\bv)
+ \tfrac{4}{7}\wp_{66}(\bu)Q_{2445}(\bv)  \big]\lambda_{3},
\\[0.5ex]
P_{23}(\bu,\bv) =
\big[ \tfrac{1}{21}\wp_{16}(\bu)Q_{4456}(\bv) - \tfrac{1}{21}Q_{3466}(\bu)\wp_{24}(\bv)
- \tfrac{1}{21}Q_{4456}(\bu)\wp_{24}(\bv) \\
\quad{}   + \tfrac{1}{21}Q_{3466}(\bu)\wp_{16}(\bv)  \big]\lambda_{6}^3
+ \big[ \tfrac{1}{3}Q_{4466}(\bu)\wp_{15}(\bv)
+ \tfrac{2}{21}\wp_{46}(\bu)Q_{3344}(\bv) - \wp_{12}(\bu)\wp_{55}(\bv)  \\
\quad{}   + \tfrac{8}{21}\wp_{16}(\bu)Q_{3466}(\bv)
+ \tfrac{1}{1512}Q_{3346}(\bu)Q_{5555}(\bv)
+ \tfrac{1}{27}Q_{4445}(\bu)\wp_{26}(\bv) + \wp_{25}(\bv)\wp_{15}(\bu) \\
\quad{}   - \tfrac{19}{84}\wp_{14}(\bu)\wp_{44}(\bv)
+ \tfrac{1}{126}Q_{3355}(\bu)\wp_{26}(\bv)
+ \tfrac{2}{21}\wp_{46}(\bv)Q_{2346}(\bu)
+ \tfrac{1}{126}Q_{3355}(\bu)\wp_{34}(\bv) \\
\quad{}   - \tfrac{1}{21}Q_{3446}(\bu)\wp_{26}(\bv)
- \tfrac{1}{21}\wp_{34}(\bu)Q_{3446}(\bv)
+ \tfrac{1}{27}\wp_{34}(\bv)Q_{4445}(\bu) + \tfrac{2}{7}\wp_{46}(\bu)Q_{2445}(\bv) \\
\quad{}   + \tfrac{8}{21}\wp_{16}(\bu)Q_{4456}(\bv)
- \tfrac{19}{756}\wp_{14}(\bu)Q_{5555}(\bv)
+ \tfrac{1}{21}\wp_{24}(\bu)Q_{4456}(\bv)   \\
\quad{}   + \tfrac{1}{21}\wp_{24}(\bu)Q_{3466}(\bv)
+ \tfrac{1}{168}Q_{3346}(\bu)\wp_{44}(\bv) \big]\lambda_{5}\lambda_{6}
+ \big[ \tfrac{8}{21}Q_{3446}(\bu)\wp_{34}(\bv) + \tfrac{1}{3}Q_{1255}(\bv) \\
\quad{}   - \tfrac{2}{9}Q_{1555}(\bu)\wp_{36}(\bv)
+ \wp_{35}(\bu)\wp_{22}(\bv)
- \tfrac{1}{3}Q_{4446}(\bv)\wp_{33}(\bu)
- \tfrac{1}{3}\wp_{66}(\bu)Q_{2246}(\bv)
\\
\quad{}
+ \tfrac{1}{6}Q_{4466}(\bu)\wp_{15}(\bv)
+ \tfrac{1}{12}Q_{3355}(\bu)\wp_{26}(\bv)
+ \tfrac{2}{3}Q_{3446}(\bu)\wp_{26}(\bv)
+ \tfrac{4}{3}\wp_{35}(\bu)\wp_{13}(\bv) \\
\quad{}
- \tfrac{1}{3}\wp_{35}(\bu)Q_{1556}(\bv)
- \tfrac{7}{9}\wp_{26}(\bu)Q_{4445}(\bv)
+ \tfrac{1}{2}Q_{5555}(\bu)\wp_{14}(\bv) + \tfrac{17}{6}\wp_{44}(\bu)\wp_{14}(\bv) \\
\quad{}   + \tfrac{1}{18}Q_{1555}(\bu)\wp_{45}(\bv)
+ \tfrac{1}{3}\wp_{66}(\bu)Q_{2344}(\bv)
- \tfrac{3}{2}Q_{4446}(\bv)\wp_{25}(\bu) - \tfrac{3}{2}\wp_{23}(\bu)\wp_{25}(\bv) \\
\quad{}   + \tfrac{3}{14}Q_{3355}(\bu)\wp_{34}(\bv)
+ \tfrac{29}{14}\wp_{46}(\bv)Q_{2445}(\bu)
+ \tfrac{2}{3}Q_{3466}(\bu)\wp_{24}(\bv) + \tfrac{9}{2}\wp_{25}(\bv)\wp_{15}(\bu) \\
\quad{}   + \tfrac{1}{3}Q_{4456}(\bu)\wp_{16}(\bv)
- \tfrac{2}{3}\wp_{34}(\bv)Q_{4445}(\bu)
+ \tfrac{1}{2}Q_{4466}(\bu)\wp_{23}(\bv) + 2\wp_{15}(\bu)\wp_{33}(\bv) \\
\quad{}   + \tfrac{1}{9}Q_{4446}(\bv)Q_{4466}(\bu)
+ \tfrac{71}{42}\wp_{46}(\bv)Q_{2346}(\bu)
- \tfrac{1}{3}\wp_{22}(\bu)Q_{5556}(\bv) \\
\quad{}   + \tfrac{2}{3}Q_{4456}(\bu)\wp_{24}(\bv)
- \tfrac{1}{3}\wp_{16}(\bu)Q_{3466}(\bv)
+ \tfrac{7}{12}\wp_{45}(\bu)Q_{2446}(\bv) + \tfrac{4}{21}\wp_{46}(\bu)Q_{3344}(\bv) \\
\quad{}   + \tfrac{1}{3}\wp_{36}(\bv)Q_{2446}(\bu)
- \tfrac{4}{9}Q_{5556}(\bu)\wp_{13}(\bv)
+ \tfrac{1}{9}Q_{1556}(\bu)Q_{5556}(\bv) + \tfrac{1}{3}Q_{1335}(\bu) \big]\lambda_{4},
 \\[0.5ex]
P_{26}(\bu,\bv) =
\big[ \tfrac{1}{2}\wp_{46}(\bu)Q_{2344}(\bv) + \tfrac{1}{252}Q_{5555}(\bu)Q_{3344}(\bv)
+ \tfrac{6}{7}\wp_{16}(\bu)Q_{3446}(\bv) \\
\quad{}   + \tfrac{1}{3}Q_{5556}(\bu)\wp_{12}(\bv)
- \tfrac{1}{9}Q_{4466}(\bu)Q_{1556}(\bv)
+ \tfrac{1}{42}\wp_{24}(\bu)Q_{3355}(\bv)  + \wp_{15}(\bu)\wp_{15}(\bv) \\
 \quad{}   + \tfrac{1}{9}\wp_{13}(\bu)Q_{4466}(\bv)
- \tfrac{1}{2}\wp_{46}(\bu)Q_{2246}(\bv)
- \tfrac{13}{42}\wp_{24}(\bu)Q_{3446}(\bv) + \tfrac{1}{4}\wp_{22}(\bu)\wp_{25}(\bv) \\
 \quad{}   - \tfrac{1}{9}\wp_{34}(\bu)Q_{1555}(\bv)
+ \tfrac{1}{2}8\wp_{44}(\bu)Q_{2346}(\bv)
+ \tfrac{3}{2}8\wp_{44}(\bu)Q_{2445}(\bv) - \tfrac{5}{3}\wp_{25}(\bu)\wp_{13}(\bv) \\
 \quad{}   + \tfrac{5}{18}\wp_{24}(\bu)Q_{4445}(\bv)
- \tfrac{1}{9}Q_{4446}(\bu)Q_{4446}(\bv)
- \tfrac{1}{12}\wp_{22}(\bv)Q_{4466}(\bu) - \wp_{35}(\bu)\wp_{12}(\bv) \\
 \quad{}   - \tfrac{1}{9}\wp_{26}(\bu)Q_{1555}(\bv)
+ \tfrac{1}{12}Q_{2446}(\bu)\wp_{26}(\bv)
+ \tfrac{1}{3}\wp_{34}(\bu)Q_{2446}(\bv)
+ \tfrac{1}{2}\wp_{66}(\bu)Q_{2244}(\bv) \\
 \quad{}   + \tfrac{1}{252}Q_{5555}(\bu)Q_{2346}(\bv)
+ \wp_{66}(\bu)Q_{1344}(\bv)
+ \tfrac{1}{28}\wp_{44}(\bu)Q_{3344}(\bv)
+ \wp_{15}(\bu)\wp_{23}(\bv) \\
 \quad{}   + \tfrac{1}{84}Q_{5555}(\bu)Q_{2445}(\bv)
+ \wp_{14}(\bu)Q_{4456}(\bv)
- \tfrac{1}{6}Q_{4446}(\bu)\wp_{23}(\bv) - \tfrac{1}{3}Q_{1556}(\bv)\wp_{25}(\bu)  \\
 \quad{}
+ \tfrac{5}{3}\wp_{14}(\bv)Q_{3466}(\bu)
- \tfrac{10}{9}\wp_{16}(\bu)Q_{4445}(\bv) + \tfrac{4}{21}\wp_{16}(\bu)Q_{3355}(\bv)  \big]\lambda_{5}
+ \big[ 6\wp_{15}(\bu)\wp_{15}(\bv) \\
 \quad{}
+ \tfrac{1}{7}Q_{3446}(\bu)\wp_{24}(\bv)
- \tfrac{1}{42}\wp_{24}(\bv)Q_{3355}(\bu)
- \tfrac{1}{126}Q_{3466}(\bu)Q_{3346}(\bv)
- \tfrac{4}{3}Q_{4446}(\bu)\wp_{15}(\bv)  \\
 \quad{}   - \tfrac{1}{7}Q_{3446}(\bu)\wp_{16}(\bv)
- \tfrac{1}{8}\wp_{66}(\bu)Q_{2244}(\bv)
+ \tfrac{1}{9}\wp_{16}(\bu)Q_{4445}(\bv)
- \tfrac{1}{4}\wp_{66}(\bu)Q_{1344}(\bv)  \\
 \quad{}   + \tfrac{19}{63}\wp_{14}(\bv)Q_{3466}(\bu)
- 3\wp_{55}(\bu)\wp_{11}(\bv)
+ \tfrac{19}{63}Q_{4456}(\bu)\wp_{14}(\bv) \\
 \quad{}   - \tfrac{1}{126}Q_{4456}(\bu)Q_{3346}(\bv)
- \tfrac{1}{9}\wp_{24}(\bu)Q_{4445}(\bv)
+ \tfrac{1}{42}Q_{3355}(\bu)\wp_{16}(\bv) \big]\lambda_{6}^2,
 \\[0.5ex]
P_{29}(\bu,\bv) =
\big[ \tfrac{1}{4}Q_{2244}(\bu)\wp_{46}(\bv)
- \tfrac{1}{54}Q_{3346}(\bv)Q_{4445}(\bu)
+ \tfrac{1}{42}Q_{3446}(\bv)Q_{3346}(\bu)  \\
 \quad{}   + \tfrac{19}{126}\wp_{14}(\bv)Q_{3355}(\bu)
+ \tfrac{1}{3}\wp_{22}(\bv)Q_{4446}(\bu)
+ \tfrac{5}{9}\wp_{13}(\bv)Q_{4446}(\bu) - \tfrac{1}{2}\wp_{55}(\bv)Q_{2234}(\bu) \\
 \quad{}   - \tfrac{1}{21}Q_{4456}(\bu)Q_{2346}(\bv)
- \tfrac{1}{21}Q_{3344}(\bv)Q_{4456}(\bu)
- \tfrac{1}{7}Q_{2445}(\bv)Q_{3466}(\bu)  \\
 \quad{}   - \tfrac{1}{252}Q_{3346}(\bv)Q_{3355}(\bu)
- \tfrac{1}{7}Q_{4456}(\bv)Q_{2445}(\bu)
- \tfrac{1}{21}Q_{2346}(\bv)Q_{3466}(\bu) \\
 \quad{}   + \tfrac{1}{3}\wp_{24}(\bv)Q_{1555}(\bu)
- \tfrac{8}{27}\wp_{14}(\bv)Q_{4445}(\bu)
- \tfrac{1}{2}Q_{2446}(\bv)\wp_{24}(\bu)
- \tfrac{1}{3}\wp_{34}(\bu)Q_{1446}(\bv) \\
 \quad{}   - \tfrac{1}{3}Q_{1555}(\bv)\wp_{16}(\bu)
- \tfrac{1}{2}\wp_{12}(\bv)Q_{4466}(\bu)
- \wp_{15}(\bv)Q_{2356}(\bu) + \tfrac{1}{6}\wp_{26}(\bv)Q_{1446}(\bu)  \\
 \quad{}   + \tfrac{23}{21}Q_{3446}(\bv)\wp_{14}(\bu)
- \tfrac{4}{3}\wp_{15}(\bv)Q_{1556}(\bu)
+ \tfrac{1}{2}Q_{1344}(\bu)\wp_{46}(\bv)
- 2\wp_{22}(\bu)\wp_{15}(\bv) \\
 \quad{}   + \tfrac{4}{9}Q_{4446}(\bu)Q_{1556}(\bv)
- \tfrac{1}{3}\wp_{55}(\bv)Q_{2226}(\bu)
+ Q_{5556}(\bv)\wp_{11}(\bu)  + Q_{2446}(\bv)\wp_{16}(\bu)  \\
\quad{}   + \tfrac{3}{2}\wp_{12}(\bv)\wp_{25}(\bu)
- 3\wp_{35}(\bv)\wp_{11}(\bu) - \tfrac{1}{21}Q_{3344}(\bv)Q_{3466}(\bu)
- \tfrac{8}{3}\wp_{15}(\bv)\wp_{13}(\bu) \big] \lambda_6,
 \\[0.5ex]
P_{32}(\bu,\bv) =
- \tfrac{1}{6}Q_{113666}(\bu)\wp_{66}(\bv)
+ \tfrac{1}{6}Q_{223466}(\bv)\wp_{46}(\bu)
- \tfrac{5}{21}Q_{2445}(\bu)Q_{3446}(\bv) \\
\quad{}
+ \tfrac{4}{3}\wp_{12}(\bv)Q_{4446}(\bu)
- \tfrac{1}{9}Q_{1555}(\bv)\wp_{14}(\bu)
- \tfrac{1}{3}Q_{3334}(\bv)\wp_{15}(\bu)
+ \tfrac{1}{3}Q_{1335}(\bu)\wp_{34}(\bv) \\
\quad{}
- \tfrac{1}{6}Q_{2446}(\bu)Q_{3445}(\bv)
+ \tfrac{1}{18}Q_{3334}(\bu)Q_{4446}(\bv)
- \tfrac{1}{42}Q_{2346}(\bu)Q_{3355}(\bv)  \\
\quad{}   - \tfrac{1}{2}\wp_{16}(\bu)Q_{2255}(\bv)
+ \tfrac{1}{2}Q_{1135}(\bu)\wp_{56}(\bv)
- \tfrac{11}{3}\wp_{16}(\bu)Q_{1446}(\bv)
+ \tfrac{3}{2}Q_{2446}(\bv)\wp_{14}(\bu) \\
\quad{}
+ \tfrac{1}{18}Q_{3344}(\bu)Q_{4445}(\bv)
+ \tfrac{1}{18}Q_{1555}(\bu)Q_{3346}(\bv)
- \tfrac{1}{6}Q_{3466}(\bu)Q_{2344}(\bv) \\
\quad{}
- \tfrac{1}{6}Q_{2226}(\bu)\wp_{35}(\bv)
+ \tfrac{1}{2}Q_{1136}(\bu)\wp_{55}(\bv)
+ \tfrac{1}{4}\wp_{25}(\bu)Q_{2334}(\bv)
+ \tfrac{2}{9}Q_{2346}(\bu)Q_{4445}(\bv)  \\
\quad{}
- \tfrac{2}{9}\wp_{13}(\bu)Q_{1556}(\bv)
- \tfrac{1}{12}Q_{2446}(\bu)Q_{3346}(\bv)
- \tfrac{1}{6}Q_{4456}(\bu)Q_{2344}(\bv)
+ \wp_{13}(\bv)\wp_{22}(\bu)
\\
\quad{}   - \tfrac{2}{9}Q_{1556}(\bu)Q_{1556}(\bv)
+ \tfrac{3}{2}Q_{1446}(\bu)\wp_{24}(\bv)
- \tfrac{1}{2}Q_{2234}(\bv)\wp_{35}(\bu)
+ \tfrac{2}{3}Q_{2356}(\bv)\wp_{13}(\bu)\\
\quad{}
- \tfrac{1}{42}Q_{3344}(\bu)Q_{3446}(\bv)
- \tfrac{4}{21}Q_{2346}(\bu)Q_{3446}(\bv)
- \tfrac{1}{12}Q_{4466}(\bu)Q_{2334}(\bv)
+ \tfrac{1}{2}Q_{1133}(\bv)  \\
\quad{}
- \tfrac{1}{2}Q_{1155}(\bu)\wp_{36}(\bv)
- \tfrac{1}{24}Q_{2244}(\bu)Q_{5555}(\bv)
- \tfrac{1}{12}Q_{1344}(\bu)Q_{5555}(\bv)
+ \wp_{33}(\bv)\wp_{11}(\bu)  \\
\quad{}   + \tfrac{1}{6}Q_{2445}(\bu)Q_{4445}(\bv)
- \tfrac{2}{3}Q_{4466}(\bu)\wp_{11}(\bv)
+ \tfrac{1}{3}\wp_{26}(\bu)Q_{1255}(\bv)
+ \tfrac{1}{3}Q_{1255}(\bu)\wp_{34}(\bv) \\
\quad{}
- \tfrac{7}{12}Q_{1344}(\bu)\wp_{44}(\bv)
+ \tfrac{1}{6}Q_{1446}(\bu)Q_{4455}(\bv)
+ \tfrac{1}{9}Q_{2226}(\bv)Q_{5556}(\bu)
- \tfrac{5}{9}\wp_{13}(\bu)\wp_{13}(\bv)  \\
\quad{}   - \tfrac{1}{14}Q_{2445}(\bu)Q_{3355}(\bv)
- \tfrac{1}{6}Q_{1335}(\bu)\wp_{26}(\bv)
- Q_{2345}(\bu)\wp_{15}(\bv)
- Q_{1235}(\bu)\wp_{45}(\bv)  \\
\quad{}    + \tfrac{1}{6}Q_{2234}(\bu)Q_{5556}(\bv)
+ \tfrac{1}{6}Q_{3466}(\bu)Q_{2246}(\bv)
- \tfrac{1}{6}Q_{4466}(\bu)Q_{2245}(\bv) - \wp_{12}(\bv)\wp_{23}(\bu)  \\
\quad{}
- \tfrac{1}{8}Q_{2244}(\bu)\wp_{44}(\bv)
+ \tfrac{1}{3}Q_{2345}(\bu)Q_{4446}(\bv)
- \tfrac{1}{42}Q_{3344}(\bu)Q_{3355}(\bv) - 4\wp_{12}(\bv)\wp_{15}(\bu) \\
\quad{}
+ \tfrac{1}{3}Q_{1556}(\bu)Q_{2356}(\bv)
- Q_{1355}(\bu)\wp_{24}(\bv)
+ Q_{2355}(\bv)\wp_{14}(\bu)
+ \wp_{16}(\bu)Q_{1355}(\bv).
\end{gather*}

\section{Solving the Jacobi inversion problem} \label{APP_JIP}

In Section \ref{SEC_KF} we explicitly solved the Jacobi inversion problem for the (3,7)-curve using equations derived from the Kleinian formula, equation~(\ref{eq:37_Klein}).  We repeated this method of solution in Section~\ref{SUBSEC_38JIP} for the (3,8)-curve and commented that this method can be easily repeated for trigonal curves of even higher genus.  We demonstrate this here.

Let $\{(x_1,y_1),\dots,(x_g,y_g)\} \in C^g$ be an arbitrary set of distinct points on a trigonal curve of genus $g$ and let $(z,w)$ be any point of this set.  Then for an arbitrary point $(x,y)$ and base point~$\infty$ on $C$ we have (see \cite{eel00}) the Kleinian formula
\[
\sum_{i,j=1}^g \wp_{ij} \left( \int_{\infty}^t \bm{du} - \sum_{k=1}^g \int_{\infty}^{x_k} \bm{du} \right)
g_i(x,y) g_j(z,w)
= \frac{F\big((x,y),(z,w)\big)}{(x-z)^2}.
\]
Here the $g_i$ are the numerators of in the basis of holomorphic dif\/ferentials and $F$ is a symmetric function calculated as the numerator of the fundamental dif\/ferential of the second kind.

For a specif\/ic curve we can use expansions of the variables in the local parameter at $\infty$ to express the formula as a series in the parameter.  Each coef\/f\/icient is a polynomial in $z$, $w$ and the $\wp$-functions that must be zero.  We label these polynomials in ascending order as $\rho_i$.  In each case, only the f\/irst two are required to solve the inversion problem.
We take the resultant of $\rho_1$ and $\rho_2$, eliminating the variable $w$, to give a new polynomial $\rho_{1,2}$ which must also equal zero.  The polynomial $\rho_1$  always has degree one in $w$, while $\rho_{1,2}$ always has degree $g$ in $z$.  We can hence solve the Jacobi inversion problem using the method of Theorems~\ref{thm:37_JIP} and~\ref{thm:38_JIP}.

We have performed these calculations for the next eight trigonal $(n,s)$-curves.  In each case the computational dif\/f\/iculty got only marginally higher and was trivial when compared to say the (4,5)-curve.  This has genus six but the corresponding calculations, detailed in \cite{MEe09}, required a much more involved method.  So it is the trigonal nature of the curves which controls the computational dif\/f\/iculty of this problem.  We present the details for the (3,10) and (3,11) curves below, and give the details for the next six online at~\cite{37web}.

\subsection{The (3,10)-curve}

This curve has genus 9 and the f\/irst two polynomials derived from the Kleinian formula are,
\begin{gather*}
\rho_1  =  - z^6 + \wp_{89}z^5 + \wp_{69}z^4 + \wp_{49}z^3 + (\wp_{99}w + \wp_{39})z^2 + (\wp_{29} + \wp_{79}w)z
+ \wp_{59}w + \wp_{19} = 0, \\
\rho_2  = ( \wp_{88} - \wp_{899} )z^5 + ( \wp_{68} - \wp_{699} )z^4 + ( \wp_{48} - 2w - \wp_{499})z^3
+ (\wp_{38} + \wp_{89}w - \wp_{999}w \\
\phantom{\rho_2  =}{}  - \wp_{399})z^2 + (\wp_{28} + \wp_{78}w - \wp_{799}w - \wp_{299})z
+ \wp_{58}w - \wp_{599}w + \wp_{18} - \wp_{199} = 0.
\end{gather*}
Taking the resultant we have
\begin{gather*}
\rho_{12} =  - 2z^9 + ( 3\wp_{89} - \wp_{999} )z^8
+ ( \wp_{78} - \wp_{799} + \wp_{89}\wp_{999} + \wp_{88}\wp_{99} - \wp_{899}\wp_{99} + 2\wp_{69} \\
\phantom{\rho_{12} =}{}   - \wp_{89}^2)z^7 + ( 2\wp_{49} - \wp_{899}\wp_{79} - \wp_{89}\wp_{78} + \wp_{89}\wp_{799}
- \wp_{69}\wp_{89} - \wp_{599} - \wp_{699}\wp_{99}  \\
\phantom{\rho_{12} =}{}+ \wp_{69}\wp_{999}  + \wp_{58} + \wp_{68}\wp_{99} + \wp_{88}\wp_{79})z^6
+ ( 2\wp_{39} - \wp_{499}\wp_{99} - \wp_{699}\wp_{79} - \wp_{899}\wp_{59}  \\
\phantom{\rho_{12} =}{}+ \wp_{89}\wp_{599}  - \wp_{89}\wp_{58} - \wp_{49}\wp_{89} - \wp_{69}\wp_{78} + \wp_{88}\wp_{59} + \wp_{69}\wp_{799}
+ \wp_{68}\wp_{79} + \wp_{48}\wp_{99}  \\
\phantom{\rho_{12} =}{} + \wp_{49}\wp_{999})z^5  + ( \wp_{68}\wp_{59} - \wp_{69}\wp_{58} + \wp_{69}\wp_{599} - \wp_{699}\wp_{59} - \wp_{499}\wp_{79}
+ \wp_{38}\wp_{99} + \wp_{39}\wp_{999}\! \! \\
\phantom{\rho_{12} =}{}   + 2\wp_{29}- \wp_{39}\wp_{89} - \wp_{399}\wp_{99} + \wp_{49}\wp_{799} + \wp_{48}\wp_{79} - \wp_{49}\wp_{78})z^4
+ (2\wp_{19} - \wp_{29}\wp_{89}  \\
\phantom{\rho_{12} =}{} + \wp_{49}\wp_{599}  - \wp_{499}\wp_{59} - \wp_{399}\wp_{79} - \wp_{49}\wp_{58} + \wp_{38}\wp_{79} - \wp_{39}\wp_{78}
+ \wp_{48}\wp_{59} - \wp_{299}\wp_{99}  \\
\phantom{\rho_{12} =}{}  + \wp_{28}\wp_{99} + \wp_{29}\wp_{999} + \wp_{39}\wp_{799})z^3
+ (\wp_{18}\wp_{99} - \wp_{299}\wp_{79} + \wp_{38}\wp_{59} - \wp_{39}\wp_{58} - \wp_{19}\wp_{89} \\
\phantom{\rho_{12} =}{}    - \wp_{199}\wp_{99}+ \wp_{29}\wp_{799} - \wp_{29}\wp_{78} + \wp_{39}\wp_{599} + \wp_{28}\wp_{79} - \wp_{399}\wp_{59}
+ \wp_{19}\wp_{999})z^2
 \\
\phantom{\rho_{12} =}{}+ (\wp_{29}\wp_{599} - \wp_{299}\wp_{59}   - \wp_{29}\wp_{58} - \wp_{199}\wp_{79} + \wp_{28}\wp_{59} + \wp_{19}\wp_{799} - \wp_{19}\wp_{78}
+ \wp_{18}\wp_{79})z  \\
\phantom{\rho_{12} =}{} + \wp_{19}\wp_{599} - \wp_{199}\wp_{59}  + \wp_{18}\wp_{59} - \wp_{19}\wp_{58} = 0.
\end{gather*}
Hence, to solve the Jacobi inversion problem we could f\/ind the nine points $(z_i,w_i)$ by calculating the nine zeros, $z_i$ of $\rho_{1,2}$ and then the corresponding points $w_i$ from the equation $\rho_{1}=0$.

\subsection{The (3,11)-curve}

This curve has genus 10 and the f\/irst two polynomials derived from the Kleinian formula are,
\begin{gather*}
\rho_1 = \wp_{10,10}z^6 + \wp_{8,10}z^5 + \wp_{6,10}z^4
+ ( \wp_{4,10} - w )z^3 + (\wp_{9,10}w + \wp_{3,10})z^2 + (\wp_{7,10}w + \wp_{2,10})z \\
\phantom{\rho_1 =}{}   + \wp_{5,10}w + \wp_{1,10} = 0, \\
\rho_2 =  - 2z^7 + ( \wp_{9,10} - \wp_{10,10,10} - \lambda_{10})z^6
+ (\wp_{8,9} - \wp_{8,10,10})z^5 + (\wp_{6,9} - \wp_{6,10,10})z^4 + (\wp_{4,9} \\
\phantom{\rho_1 =}{}   - \wp_{4,10,10})z^3 + (\wp_{9,9}w + \wp_{3,9} - \wp_{9,10,10}w - \wp_{3,10,10})z^2
+ (\wp_{7,9}w + \wp_{2,9} - \wp_{7,10,10}w \\
\phantom{\rho_1 =}{}   - \wp_{2,10,10})z + \wp_{1,9} - \wp_{1,10,10} - \wp_{5,10,10}w + \wp_{5,9}w = 0.
\end{gather*}
Taking the resultant we have
\begin{gather*}
\rho_{12} = 2z^{10} + (\lambda_{10} - 3\wp_{9,10} + \wp_{10,10,10})z^9
+ ( \wp_{10,10}\wp_{9,10,10} - \wp_{8,9} - \wp_{9,10}\lambda_{10} - \wp_{10,10,10}\wp_{9,10} \\
\phantom{\rho_{12} =}{}   - 2\wp_{7,10} + \wp_{8,10,10} + \wp_{9,10}^2 - \wp_{10,10}\wp_{9,9})z^8
+ (\wp_{8,10}\wp_{9,10,10} - \lambda_{10}\wp_{7,10} - \wp_{10,10}\wp_{7,9} \\
\phantom{\rho_{12} =}{}   + \wp_{9,10}\wp_{7,10} - 2\wp_{5,10} + \wp_{8,9}\wp_{9,10} - \wp_{8,10}\wp_{9,9}
- \wp_{10,10,10}\wp_{7,10} - \wp_{8,10,10}\wp_{9,10} + \wp_{6,10,10} \\
\phantom{\rho_{12} =}{}   + \wp_{10,10}\wp_{7,10,10} - \wp_{6,9})z^7
+ ( \wp_{6,9}\wp_{9,10} - \wp_{8,10}\wp_{7,9}  - \wp_{10,10}\wp_{5,9} - \lambda_{10}\wp_{5,10}
 \\
\phantom{\rho_{12} =}{} + \wp_{10,10}\wp_{5,10,10}  - \wp_{6,10}\wp_{9,9} - \wp_{8,10,10}\wp_{7,10} + \wp_{9,10}\wp_{5,10} + \wp_{6,10}\wp_{9,10,10}
+ \wp_{8,9}\wp_{7,10} \\
\phantom{\rho_{12} =}{} + \wp_{8,10}\wp_{7,10,10}   - \wp_{6,10,10}\wp_{9,10} - \wp_{4,9} - \wp_{10,10,10}\wp_{5,10} + \wp_{4,10,10})z^6
+ ( \wp_{4,10}\wp_{9,10,10}  \\
\phantom{\rho_{12} =}{}- \wp_{4,10}\wp_{9,9} + \wp_{8,9}\wp_{5,10}   - \wp_{4,10,10}\wp_{9,10} + \wp_{8,10}\wp_{5,10,10} + \wp_{3,10,10} - \wp_{8,10}\wp_{5,9}
 \\
\phantom{\rho_{12} =}{}- \wp_{8,10,10}\wp_{5,10} + \wp_{4,9}\wp_{9,10} + \wp_{6,9}\wp_{7,10}   + \wp_{6,10}\wp_{7,10,10} - \wp_{6,10}\wp_{7,9} - \wp_{6,10,10}\wp_{7,10}  \\
\phantom{\rho_{12} =}{} - \wp_{3,9})z^5
+ (\wp_{6,9}\wp_{5,10} - \wp_{2,9} + \wp_{2,10,10}  - \wp_{6,10,10}\wp_{5,10} - \wp_{3,10,10}\wp_{9,10} - \wp_{4,10,10}\wp_{7,10}
 \\
\phantom{\rho_{12} =}{} + \wp_{6,10}\wp_{5,10,10} + \wp_{3,9}\wp_{9,10} + \wp_{4,9}\wp_{7,10}  - \wp_{4,10}\wp_{7,9} + \wp_{3,10}\wp_{9,10,10} - \wp_{6,10}\wp_{5,9}  \\
\phantom{\rho_{12} =}{} + \wp_{4,10}\wp_{7,10,10}
- \wp_{3,10}\wp_{9,9})z^4 + (\wp_{1,10,10}  - \wp_{4,10,10}\wp_{5,10} + \wp_{3,9}\wp_{7,10} - \wp_{2,10,10}\wp_{9,10} \\
\phantom{\rho_{12} =}{}  + \wp_{2,9}\wp_{9,10}
+ \wp_{4,9}\wp_{5,10} + \wp_{4,10}\wp_{5,10,10}  - \wp_{3,10,10}\wp_{7,10} - \wp_{2,10}\wp_{9,9} + \wp_{2,10}\wp_{9,10,10} \\
\phantom{\rho_{12} =}{} + \wp_{3,10}\wp_{7,10,10} - \wp_{4,10}\wp_{5,9} - \wp_{3,10}\wp_{7,9} - \wp_{1,9})z^3   + (\wp_{1,10}\wp_{9,10,10} - \wp_{2,10,10}\wp_{7,10}  \\
\phantom{\rho_{12} =}{} - \wp_{1,10}\wp_{9,9} - \wp_{2,10}\wp_{7,9}
+ \wp_{2,10}\wp_{7,10,10} + \wp_{3,9}\wp_{5,10}  + \wp_{1,9}\wp_{9,10} - \wp_{3,10,10}\wp_{5,10}  \\
\phantom{\rho_{12} =}{}  - \wp_{1,10,10}\wp_{9,10} + \wp_{3,10}\wp_{5,10,10}
+ \wp_{2,9}\wp_{7,10} - \wp_{3,10}\wp_{5,9})z^2 + ( \wp_{1,10}\wp_{7,10,10} - \wp_{2,10}\wp_{5,9} \\
\phantom{\rho_{12} =}{}   - \wp_{1,10,10}\wp_{7,10} + \wp_{1,9}\wp_{7,10}
+ \wp_{2,10}\wp_{5,10,10} - \wp_{1,10}\wp_{7,9} - \wp_{2,10,10}\wp_{5,10} + \wp_{2,9}\wp_{5,10})z\\
\phantom{\rho_{12} =}{}
+ \wp_{1,9}\wp_{5,10} + \wp_{1,10}\wp_{5,10,10} - \wp_{1,10}\wp_{5,9} - \wp_{1,10,10}\wp_{5,10} = 0.
\end{gather*}
Hence, to solve the Jacobi inversion problem we could f\/ind the ten points $(z_i,w_i)$ by calculating the ten zeros, $z_i$ of $\rho_{1,2}$ and then the corresponding points $w_i$ from the equation $\rho_{1}=0$.

\subsection*{Acknowledgments}

Many thanks to Professor Chris Eilbeck for useful conversations.  Thanks also to the three anonymous referees who gave constructive and useful comments.

\pdfbookmark[1]{References}{ref}
\LastPageEnding

\end{document}